\documentclass[fleqn,10pt]{wlscirep}
\usepackage[utf8]{inputenc}
\usepackage[T1]{fontenc}
\usepackage{multirow}
\usepackage{ulem} 
\title{Online Gambling of Pure Chance: Wager Distribution, Risk Attitude, and Anomalous Diffusion}

\author[1,2]{Xiangwen Wang}
\author[1,2,3,*]{Michel Pleimling}
\affil[1]{Department of Physics, Virginia Tech, Blacksburg, VA 24061-0435, USA}
\affil[2]{Center for Soft Matter and Biological Physics, Virginia Tech, Blacksburg, VA 24061-0435, USA}
\affil[3]{Academy of Integrated Science, Virginia Tech, Blacksburg, VA 24061-0563, USA}

\affil[*]{pleim@vt.edu}

\begin{abstract}
Online gambling sites offer many different gambling games. In this work we analyse the gambling logs of numerous
solely probability-based gambling games and extract the wager and odds distributions. We find that the log-normal
distribution describes the wager distribution at the aggregate level. Viewing the gamblers' net incomes as 
random walks, we study the mean-squared displacement of net income 
and related quantities and find different diffusive behaviors for
different games. We discuss possible origins for the observed anomalous diffusion.
\end{abstract}
\begin{document}

\flushbottom
\maketitle
%
%
\thispagestyle{empty}

\section*{Introduction}

Today, gambling is a huge industry with a huge social impact.
According to a report by the American Gaming Association~\cite{AGA2018},
commercial casinos in the United States alone made total revenue of over 40
billion US dollars in 2017. On the other hand, different studies reported
that $0.12\%-5.8\%$ of the adults and $0.2\%-12.3\%$ of the adolescents
across different countries in the world are
experiencing problematic gambling~\cite{Calado2016, Calado2017}.
Studying the gamblers' behavior patterns not only contributes to the
prevention of problematic gambling and adolescent gambling,
but also helps to better understand human decision-making processes.
Researchers have put a lot of attention on studying
gambling-related activities.
Economists have proposed many theories about how humans make decisions under
different risk conditions. Several of them can also be applied to model
gambling behaviors. For example, the prospect theory introduced by
Kahneman and Tversky~\cite{Kahneman1979} and its variant cumulative
prospect theory~\cite{Tversky1992} have been adopted in modeling
casino gambling~\cite{Barberis2012}.
In parallel to the theoretical approach, numerous studies focus on
the empirical analysis of gambling behaviors, aiming at explaining the
motivations behind problematic gambling behaviors.
However, parametric models that quantitatively describe empirical
gambling behaviors are still missing.
Such models can contribute to evaluating gambling theories proposed by
economists, as well as yield a better understanding of the gamblers' behaviors.
Our goal is to provide such a parametric model for
describing human wagering activities and risk attitude during gambling from
empirical gambling logs. However, it is very difficult to obtain gambling logs
from traditional casinos, and it is hard to collect large amounts of
behavior data in a lab-controlled environment. Therefore in this paper
we will focus on analyzing online gambling logs collected from
online casinos.

Whereas historically the development of probability theory, which then became
the foundation of statistics, was tied to chance games,
today we use statistical tools to analyze gamblers' behaviors.

Recent years have seen an increasing trend of online gambling due to its
low barriers to entry, high anonymity and instant payout.
For researchers of gambling
behaviors, online gambling games present two advantages:
simple rules and the availability of large amounts of gambling logs.
In addition to the usual forms of gambling games that can be found in
traditional casinos, many online casinos also offer games that follow
very simple rules, which makes analyzing the gambling behavior much
easier as there are much fewer degrees of freedom required to be considered.
On the other hand, many online casinos have made gambling
logs publicly available on their websites, mainly for
verification purposes, which provides researchers with abundant data to work on.
Due to the high popularity of online gambling, in a dataset
provided by an online casino there are often thousands or even
hundreds of thousands of gamblers listed. Such a large scale of data can
hardly be obtained in a lab environment. Prior research has begun to
make use of online gambling logs. For example,
Meng's thesis~\cite{Meng2018} presented a pattern analysis of typical
gamblers in Bitcoin gambling.
It is worth arguing that although our work only focuses on
the behaviors of online gamblers, there is no reason to think that
our conclusions cannot be extended to traditional gamblers.

Naturally, we can treat the changing cumulative net income of a
player during their gambling activities as a random walk process \cite{Wang2018}.
We are particularly interested in the diffusive characteristics of
the gambler's net income. This is another reason why we want to analyze
the wager distribution and risk attitude of gamblers,
since both distributions are closely related to the displacement
distribution for the gambler's random walks. Within this paper, we will
mainly focus on the analysis at the population level.
Physicists have long been studying diffusion processes in different
systems, and recently anomalous diffusive properties have been reported
in many human activities, including human spatial movement~\cite{Rhee2011,Brockmann2008b, Kim2010}, 
and information foraging~\cite{Wang2017}.
In a previous study of skin gambling~\cite{Wang2018},
we have shown that in a parimutuel betting game (where players gamble against
each other), a gambler's net income displays a crossover from superdiffusion
to normal diffusion. We have reproduced this crossover in simulations by
introducing finite and overall conserved gamblers' wealth (see \cite{Toscani2019}
for a different way of modeling this using kinetic equations of Boltzmann and
Fokker-Planck type).
However, this explanation cannot be used in other types of gambling games
where there is no interaction among gamblers
(e.g., fixed-odds betting games, which will be introduced below),
as they violate the conservation of gamblers' overall wealth.
In this paper, we want to expand the scope of our study to
more general gambling games, check the corresponding diffusive properties,
and propose some explanations for the observed behaviors.

One of our goals is to uncover the commonalities behind the behavior of online gamblers.
To implement this,
we analyze the data from different online gambling systems.
The first one is skin gambling, where the bettors are mostly video game
players and where cosmetic skins from online video games are used as virtual
currency for wagering~\cite{Wang2018, Holden2018}.
The other system is crypto-currency gambling,
where the bettors are mostly crypto-currency users. Different types
of crypto-currencies are used for wagering. Commonly used crypto-currencies 
include Bitcoin, Ethereum, and Bitcoin Cash, whose basic units are 
BTC, ETH and BCH, respectively.
As the overlap of these two communities, video game players and
crypto-currency users, is relatively small for now,
features of gambling patterns common between these two gambling systems
are possibly features common among all online gamblers.

Not only do we consider different gambling systems,
but we also discuss different types of gambling games.
In this paper, we discuss four types of solely probability-based
gambling games (Roulette, Crash, Satoshi Dice and Jackpot),
whose outcomes in theory will
not benefit from the gamblers' skill or experience
when the in-game random number generators are well designed.
In general, there are two frameworks of betting in gambling:
fixed-odds betting,
where the odds is fixed and known before players wager in one round;
and parimutuel betting, where the odds can still change after players
place the bets until all players finish wagering.
In fixed-odds betting, usually players bet against the house/website,
and there is no direct interaction among players; and in
parimutuel betting, usually players bet against each other.
The four types of games we discuss in this paper will
cover both betting frameworks (see the Methods section).

When a player attends one round in any of those games,
there are only two possible outcomes: either win or lose.
When losing, the player will lose the wager they placed during that round;
whereas when winning, the prize winner receives equals
their original wager multiplied by a coefficient.
This coefficient is generally larger than $1$, and
in gambling terminology, it is called odds in decimal
format~\cite{Buhagiar2018, Rodriguez2017}.
Here we will simply refer to it as odds.
Note that the definition of odds in gambling is different than the definition
of odds in statistics, and in this paper we follow the former one.
When a player attends one round, their chance of winning is usually close to,
but less than the inverse of the odds.
The difference is caused by the players' statistical disadvantage
in winning compared to the house due to the design of the game rules.
In addition, the website usually charges the winner with a site cut
(commission fee), which is a fixed percentage of the prize.

We further define the {\it payoff}, $o_p$, to be the net change of
one player's wealth after they attend one round.
Although the four types of games are based on different rules,
the payoffs all follow the same expression
\begin{equation}\label{eq:payoff}
\displaystyle o_p = \left\lbrace \begin{split}
\displaystyle &-b, &\text{with\ probability\ } &p= 1- \frac{1}{m} + f_\text{m}, \\
\displaystyle &(1-\eta)(m-1) b, &\text{with\ probability\ } &q=1-p = \frac{1}{m} - f_\text{m},
\end{split}\right.
\end{equation}
where $b>0$ is the wager the player places, $m>1$ is the odds,
$1>\eta\geq 0$ corresponds to the site cut, and $f_\text{m}$
is a non-negative value based on the odds representing
the players' statistical disadvantage in winning, as mentioned earlier.
At least either $\eta$ or $f_\text{m}$ are non-zero.

From Eq.~(\ref{eq:payoff}), we can obtain the expected payoff of attending one round
\begin{equation}
\label{eq:expected-payoff}
\displaystyle E\left( o_p|m, b\right)
= \Big(- (1-1/m +f_m) + (1-\eta)(m-1)(1/m - f_m)\Big)b
= - \Big((1-\eta)m f_m + (1-1/m + f_m)\eta \Big)b\equiv -\xi b,
\end{equation}
which is always negative since either $\eta$ or $f_\text{m}$ are non-zero.
In gambling terminology, $\xi$ is called the house edge,
from which the websites make profits.
The house edge represents the proportion the website will benefit on
average when players wager. In the four types of games we discuss,
the house edge $\xi$ ranges from $1\%$ to $8\%$.
If there is no house edge $\xi=0$, that means it is a fair game.
In a fair game or when we ignore the house edge, the expected payoff would be 0.

In the Results section, we begin with an analysis of wager distribution and log-ratios 
between successive wagers, which helps us to understand the gamblers' wagering strategy. 
We then focus on an analysis of risk attitude by studying the distribution of the odds players choose to wager with. 
We conclude by extending our discussion to the analysis of net incomes of gamblers viewed as random walks.
This allows us to gain insights into the gamblers' behaviors by computing quantities like
the ensemble/time-averaged mean-squared displacement, the first-passage time distribution, ergodicity
breaking parameter, and Gaussianity. 
Detailed information about the games and datasets discussed in this paper can be found in the Methods section.

\section*{Results}

\subsection*{Wager distribution}
From the viewpoint of the interaction among players,
the games discussed in this paper can be grouped into
two classes: in Roulette, Crash, and Satoshi Dice games,
there is little or no interaction among players,
whereas in Jackpot games, players need to gamble against each other.
At the same time, from the viewpoint of wager itself,
the games can also be grouped into two classes:
In games (A-G), the wagers can be an arbitrary amount of virtual currencies,
such as virtual skin tickets or crypto-currency units, whereas in game (H),
the wagers are placed in the form of in-game skins,
which means the wager distribution further involves the distributions of
the market price and availability of the skins.

Furthermore, from the viewpoint of the odds,
considering the empirical datasets we have,
when analyzing the wager distribution, there are three situations:
i) For Roulette and Satoshi Dice games, the odds are fixed constants,
and wagers placed with the same odds are analyzed to find the distribution.
ii) For Crash games, the odds are selected by the players, and wagers
placed with different odds are mixed together during distribution analysis.
iii) For the Jackpot game, the odds are not fixed at the moment when the player wagers.

\begin{table}[!ht]
\centering
\caption{The best-fitted distribution and estimated parameters of wagers.
For games (A, B, C, E, F, G) the best-fitted model is a log-normal distribution,
and for game (D) the log-normal distribution is truncated at a maximum value.
For game (H) the wager distribution follows a power law - exponential -
power law pattern. 
In the rightmost column, $\mu$ (respectively $\sigma^2$) represents the mean
(respectively variance) of the logarithms of bet values.}
\resizebox{\textwidth}{!}{
\begin{tabular}{|c|c|c|c|c|c|c|c|}
\hline
\begin{tabular}[c]{@{}c@{}}Game\\ Name\end{tabular}
 & \begin{tabular}[c]{@{}c@{}}Game\\ Category\end{tabular}
 & \begin{tabular}[c]{@{}c@{}}Wager\\ Currency\end{tabular}
 & \begin{tabular}[c]{@{}c@{}}Arbitrary\\ Bet\end{tabular}
 & Max Bet
 & Odds
 & \begin{tabular}[c]{@{}c@{}}Best-Fitted\\ Model\end{tabular}
 & Parameters                                                                                                                                                                   \\ \hline
\multirow{6}{*}{\begin{tabular}[c]{@{}c@{}}csgofast-\\ Double\\ (A)\end{tabular}}
 & \multirow{10}{*}{Roulette}
 & \multirow{12}{*}{\begin{tabular}[c]{@{}c@{}}Virtual\\ Skin\\ Ticket\end{tabular}}
 & \multirow{16}{*}{\textbf{Yes}}
 & \multirow{3}{*}{\begin{tabular}[c]{@{}c@{}}$500,000$\\ $(A_1)$\end{tabular}}
 & \begin{tabular}[c]{@{}c@{}}$2$\\ (Red)\end{tabular}
 & \multirow{16}{*}{\textbf{Log-normal}}
 & \begin{tabular}[c]{@{}c@{}}$\mu=3.689$, $\sigma=1.952$\\ $x_{min}=21$\end{tabular}                                                                                          \\ \cline{6-6} \cline{8-8}
 & & & &
 & \begin{tabular}[c]{@{}c@{}}$2$\\ (Black)\end{tabular}
 &
 & \begin{tabular}[c]{@{}c@{}}$\mu=3.807$, $\sigma=1.922$\\ $x_{min}=21$\end{tabular}                                                                                          \\ \cline{6-6} \cline{8-8}
 & & & & & \begin{tabular}[c]{@{}c@{}}$14$\\ (Green)\end{tabular}
 &
 & \begin{tabular}[c]{@{}c@{}}$\mu=3.972$, $\sigma=1.647$\\ $x_{min}=21$\end{tabular}                                                                                          \\ \cline{5-6} \cline{8-8}
 & & &
 & \multirow{3}{*}{\begin{tabular}[c]{@{}c@{}}$50,000$\\ $(A_2)$\end{tabular}}
 & \begin{tabular}[c]{@{}c@{}}$2$\\ (Red)\end{tabular}
 &
 & \begin{tabular}[c]{@{}c@{}}$\mu=2.936$, $\sigma=2.108$\\ $x_{min}=11$\end{tabular}                                                                                          \\ \cline{6-6} \cline{8-8}
 & & & &
 & \begin{tabular}[c]{@{}c@{}}$2$\\ (Black)\end{tabular}
 &
 & \begin{tabular}[c]{@{}c@{}}$\mu=3.175$, $\sigma=2.118$\\ $x_{min}=12$\end{tabular}                                                                                          \\ \cline{6-6} \cline{8-8}
 & & & &
 & \begin{tabular}[c]{@{}c@{}}$14$\\ (Green)\end{tabular}
 &
 & \begin{tabular}[c]{@{}c@{}}$\mu=2.633$, $\sigma=2.113$\\ $x_{min}=14$\end{tabular}                                                                                          \\ \cline{1-1} \cline{5-6} \cline{8-8}
 \multirow{4}{*}{\begin{tabular}[c]{@{}c@{}}csgofast-\\ X50\\ (B)\end{tabular}}
 & & &
 & \multirow{4}{*}{$50,000$}
 & \begin{tabular}[c]{@{}c@{}}$2$\\ (Blue)\end{tabular}
 &
 & \begin{tabular}[c]{@{}c@{}}$\mu=2.734$, $\sigma=1.930$\\ $x_{min}=11$\end{tabular}                                                                                          \\ \cline{6-6} \cline{8-8}
 & & & &
 & \begin{tabular}[c]{@{}c@{}}$3$\\ (Red)\end{tabular}
 &
 & \begin{tabular}[c]{@{}c@{}}$\mu=2.450$, $\sigma=2.030$\\ $x_{min}=12$\end{tabular}                                                                                          \\ \cline{6-6} \cline{8-8}
 & & & &
 & \begin{tabular}[c]{@{}c@{}}$5$\\ (Green)\end{tabular}
 &
 & \begin{tabular}[c]{@{}c@{}}$\mu=2.814$, $\sigma=1.999$\\ $x_{min}=12$\end{tabular}                                                                                          \\ \cline{6-6} \cline{8-8}
 & & & &
 & \begin{tabular}[c]{@{}c@{}}$50$\\ (Gold)\end{tabular}
 &
 & \begin{tabular}[c]{@{}c@{}}$\mu=3.416$, $\sigma=1.548$\\ $x_{min}=11$\end{tabular}                                                                                          \\ \cline{1-2} \cline{5-6} \cline{8-8}
 \multirow{2}{*}{\begin{tabular}[c]{@{}c@{}}csgofast-\\ Crash\\ (C)\end{tabular}}
 & \multirow{3}{*}{Crash}
 & &
 & \begin{tabular}[c]{@{}c@{}}$10,000$\\ $(C_1)$\end{tabular}
 & \multirow{3}{*}{\begin{tabular}[c]{@{}c@{}}Player-\\ Selected\end{tabular}}
 &
 & \begin{tabular}[c]{@{}c@{}}$\mu=1.647$, $\sigma=2.226$\\ $x_{min}=15$\end{tabular}                                                                                          \\ \cline{5-5} \cline{8-8}
 & & &
 & \begin{tabular}[c]{@{}c@{}}$20,000$\\ $(C_2)$\end{tabular}
 & &
 & \begin{tabular}[c]{@{}c@{}}$\mu=1.932$, $\sigma=2.143$\\ $x_{min}=11$\end{tabular}                                                                                          \\ \cline{1-1} \cline{3-3} \cline{5-5} \cline{8-8}
 \begin{tabular}[c]{@{}c@{}}Ethcrash\\ (D)\end{tabular}
 &
 & \multirow{3}{*}{\begin{tabular}[c]{@{}c@{}}Crypto-\\ currency\end{tabular}}
 &
 & $0.25$ ETH
 & &
 & \begin{tabular}[c]{@{}c@{}}$\mu=-7.186$, $\sigma=6.356$\\ $x_{min}=1$\end{tabular}                                                                                          \\ \cline{1-2} \cline{5-6} \cline{8-8}
 \begin{tabular}[c]{@{}c@{}}Satoshi\\ dice\\ (E)\end{tabular}
 & \multirow{2}{*}{\begin{tabular}[c]{@{}c@{}}Satoshi\\ Dice\end{tabular}}
 & &
 & $10$ BCH
 & \multirow{2}{*}{$1.98$}
 &
 & \begin{tabular}[c]{@{}c@{}}$\mu=5.910$, $\sigma=2.691$\\ $x_{min}=34$\end{tabular}                                                                                          \\ \cline{1-1} \cline{5-5} \cline{8-8}
 \begin{tabular}[c]{@{}c@{}}Coinroll\\ (F)\end{tabular}
 & & &
 & $3$ BTC
 & &
 & \begin{tabular}[c]{@{}c@{}}$\mu=1.930$, $\sigma=2.638$\\ $x_{min}=2$\end{tabular}                                                                                           \\ \cline{1-3} \cline{5-6} \cline{8-8}
 \begin{tabular}[c]{@{}c@{}}csgospeed\\(G)\end{tabular}
 & \multirow{2}{*}{Jackpot}
 & \begin{tabular}[c]{@{}c@{}}Virtual \\ Skin\\ Ticket\end{tabular}
 &
 & $500,000$
 & \multirow{2}{*}{Not-fixed}
 &
 & \begin{tabular}[c]{@{}c@{}}$\mu=5.167$, $\sigma=1.301$\\ $x_{min}=23$\end{tabular}                                                                                          \\ \cline{1-1} \cline{3-5} \cline{7-8}
 \begin{tabular}[c]{@{}c@{}}csgofast-\\ jackpot\\ (H)\end{tabular}
 &
 & \begin{tabular}[c]{@{}c@{}}In-game\\ Skin\end{tabular}
 & \textbf{No}
 & \begin{tabular}[c]{@{}c@{}}$15$ items\\ $180,000$\\ per item\end{tabular}
 &
 & \begin{tabular}[c]{@{}c@{}}Power Law -\\ Exponential -\\ Power Law \end{tabular}
 & \begin{tabular}[c]{@{}c@{}}$\alpha=0.802$, $\delta=2.457\times10^2$\\
                              $\beta=7.080\times10^{-3}$, $\eta=3.783$\\
                              $\lambda=8.625\times10^{-5}$, $x_{min}=250$\end{tabular}
\\ \hline
\end{tabular}%
}
\label{table:wager-distribution}
\end{table}

\begin{figure}[!ht]
\includegraphics[width=\columnwidth,clip=true]{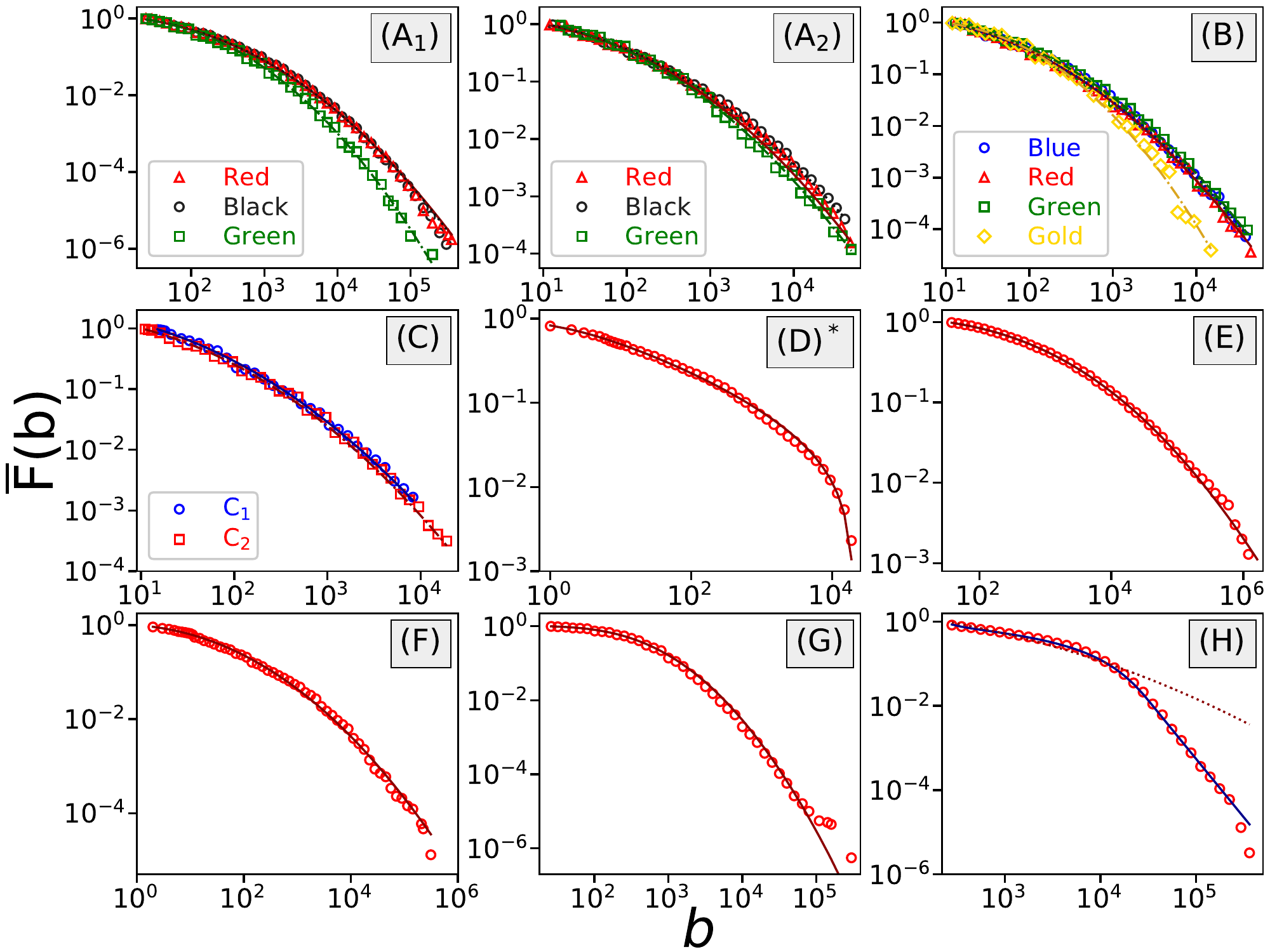}
\caption{
\label{fig:wager_distribution}
In games (A-G), where players are allowed to choose arbitrary bet values,
the wager distribution can be best fitted by
log-normal distributions~(\ref{eq:log-normal}). In game (D),
the log-normal distribution is truncated at its maximum bet value,
indicated by $^*$. The fitting lines represent the log-normal fittings.
Wagers placed under the different maximum allowed bet values are discussed separately,
e.g., in game (A), (A$_1$) has a maximum bet value of $500,000$, and (A$_2$) has
a maximum bet value of $50,000$.
On the other hand, in game (H) where wagers can only be in-game skins,
the wager distribution is best described by a pairwise power law with an
exponential transition, see Eq.~(\ref{eq:powerlaw_exp_powerlaw}). The red dotted line
represents the log-normal fitting and the blue solid line represents the
fitting of a pairwise power law with an exponential transition.}
\end{figure}

In Table~\ref{table:wager-distribution} we categorize the 8 datasets
based on the above information. At the same time, for each dataset
we perform a distribution analysis of wagers at the aggregate level.
Within the same dataset wagers placed under different
maximum allowed bet values are discussed separately.
We plot the complementary cumulative distribution function (CCDF)
of the empirical data and the fitted distribution to check the
goodness-of-fit, see Fig.~\ref{fig:wager_distribution}.
CCDF, sometimes also referred to as the survival function, is given by
$\bar{F}(x) = P(X>x) = 1 - P(X\le x)$.

It turns out that when players are allowed to place arbitrary wagers
(games A-G in Table~\ref{table:wager-distribution}), the wager
distributions can in general be best-fitted by log-normal distributions.
In particular, in games (A, B, C, E, F, G), the wager distribution can be
approximated by the following expression
\begin{equation}\label{eq:log-normal}
\displaystyle P(x) =
\frac{\displaystyle\Phi\left(\frac{\ln(x+1)-\mu}
{\sigma}\right)-\Phi\left(\frac{\ln(x)-\mu}{\sigma}\right)}
{\displaystyle 1-\Phi\left(\frac{\ln(x_\text{min})-\mu}{\sigma}\right)},
\end{equation}
with $x_\text{min}\le x$ and $\sigma>0$. $\Phi(\cdot)$
is the cumulative distribution function of the standard normal distribution.
Meanwhile in game (D), the fitted log-normal distribution
is truncated at an upper boundary $x_\text{max}$, which might result
from the maximum allowed small bet value and the huge variation of the
market price of crypto-currencies.

During model selection, we notice that when we select
different $x_\text{min}$, occasionally a power-law distribution
with exponential cutoff is reported to be a better fit, but often
it does not provide a decent absolute fit on the tail, and overall
the log-normal distribution provides smaller Kolmogorov-Smirnov distances,
see the Methods section.

On the other hand, as we have pointed out in the previous study~\cite{Wang2018},
when players are restricted to use in-game skins as wagers for gambling,
the wager distribution can be best fitted by
a shifted power law with exponential cutoff.
Now, with a similar situation in game (H), where wagers can only be in-game skins,
we find that the early part of the curve can be again fitted by
a power law with exponential cutoff, as shown in Fig.~\ref{fig:wager_distribution}(H).
However, this time it does not maintain the
exponential decay of its tail; instead, it changes back to a power-law decay.
The overall distribution contains six parameters, given by the expression

\begin{equation}
\label{eq:powerlaw_exp_powerlaw}
\displaystyle P(x) = \left\lbrace \begin{split}
\displaystyle
&\frac{1}{c_1+c_2 c_3}\ 
\frac{(x-\delta)^{-\alpha}}{1+e^{\lambda(x-\beta)}},\ 
&\text{for}\ x \le x_\text{trans}, \\
\displaystyle
&\frac{c_3}{c_1+c_2 c_3}\ x^{-\eta}, \ 
&\text{for}\ x > x_\text{trans},
\end{split}\right.
\end{equation}
where
$\displaystyle c_1=\sum\limits_{x=x_\text{min}}^{x_\text{trans}}
\frac{(x-\delta)^{-\alpha}}{1+e^{\lambda(x-\beta)}}$,
$\displaystyle c_2=\zeta\left(\eta, x_\text{trans}\right)$,
and $\displaystyle c_3=x_\text{trans}^\eta
\frac{\left(x_\text{trans}-\delta\right)^{-\alpha}}
{1+e^{\lambda\left(x_\text{trans}-\beta\right)}}$.

We believe that when players are restricted to use in-game skins as
wagers, the decision to include one particular skin in their wager
is further influenced by the price and availability of that skin.
These factors make the wager distribution deviate from the
log-normal distribution, which is observed in games (A-G).
This is very clear when comparing the wager distributions of games
(G) and (H) as both games are jackpot games of skin gambling, and the
only difference is whether players are directly using skins as wagers
or are using virtual skin tickets obtained from depositing skins.
The power-law tail, which was not observed in the previous study~\cite{Wang2018},
might result from the increment of the maximum allowed skin price
(from \$$400$ to \$$1 800$).

The above discussions, including the results for games (A-G) in
Table~\ref{table:wager-distribution}, show that the wager distributions
in pure probability-based gambling games, no matter whether the game follows
parimutuel betting or fixed-odds (preset/player-selected) betting,
stay log-normal as long as the players are allowed to place
arbitrary amounts of wagers. This commonality of log-normal distribution
no longer holds when this arbitrariness of wager value is violated,
e.g., in the scenario where the player can only wager items (in-game skins).

Log-normal distribution has been reported in a wide range of economic, biological,
and sociological systems~\cite{LIMPERT2001}, including income, species abundance,
family size, etc.
Economists have proposed different kinds of generative mechanisms for log-normal
distributions (and power-law distributions as well).
One particular interest for us is the multiplicative
process~\cite{Gibrat1930, mitzenmacher2003}.
Starting from an initial value $X_0$, random variables in a multiplicative
process follow an iterative formula
$X_{i+1}=\exp(\nu_i) X_i$ or $\ln X_{i+1}= \ln X_i + \nu_i.$
If the $v_i$ has finite mean and variance, and is independent and identically
distributed, then according to the central limit theorem, for large $i$,
$\ln X_i$ will follow a normal distribution, which means
$X_i$ will follow a log-normal distribution.

If we want to check whether gamblers follow multiplicative processes when they wager,
we can first check the correlation between consecutive
bets $(b_i, b_{i+1})$. Due to the large variances of the wager
distributions, Pearson's correlation coefficient may perform poorly.
Instead, we adopt two rank-based correlation coefficients,
Kendall's Tau \cite{Kendall1990} $\tau_K$ and Spearman's Rho \cite{Taylor1987} $\rho_S$.
At the same time, we also check the mean and variance of the log-ratios
$\ln(b_{i+1}/b_{i})$ between consecutive bets. These statistics
can be found in Table~\ref{table:log_ratio}. The results reveal that the
values of consecutive bets exhibit a strong positive correlation, with all
the correlation coefficients larger than $0.5$. It shows that players'
next bet values are largely dependent on their previous bet values.
At the same time, the bet values are following gradual changes, rather than
rapid changes. These conclusions can be confirmed by the small
mean values and small variances of log-ratios between consecutive bets.

\begin{table}[ht!]
\centering
\caption{Correlation analysis shows that there is a strong positive correlation
between consecutive bets, along with the small mean values and variances of
log-ratio between consecutive bets. Satoshi Dice (E) is
excluded here as individual gamblers in the dataset are not distinguishable.
csgofast-Jackpot (H) is excluded in the calculation of $P(b_i=b_{i+1})$
due to the low precision of bet values in this dataset.}
\resizebox{\textwidth}{!}{
\begin{tabular}{|c|c|c|c|c|c|}
\hline
Dataset & $\tau_K(b_i, b_{i+1})$ & $\rho_s(b_i, b_{i+1})$ &
$\langle\log_{10}(b_{i+1}/b_i)\rangle$ &
$\text{var}(\log_{10}(b_{i+1}/b_i))$ &
$P(b_i=b_{i+1})$ \\ \hline
csgofast-Double (A)   & $0.596$ & $0.737$ & $0.010$ & $0.183$ & $0.342$ \\ \hline
csgofast-X50 (B)      & $0.692$ & $0.803$ & $0.007$ & $0.102$ & $0.512$ \\ \hline
csgofast-Crash (C)    & $0.858$ & $0.909$ & $0.004$ & $0.038$ & $0.802$ \\ \hline
ethCrash (D)          & $0.866$ & $0.949$ & $0.000$ & $0.147$ & $0.549$ \\ \hline
Coinroll (F)          & $0.826$ & $0.925$ & $0.000$ & $0.282$ & $0.497$ \\ \hline
csgospeed-Jackpot (G) & $0.522$ & $0.675$ & $0.002$ & $0.288$ & $0.136$ \\ \hline
csgofast-Jackpot (H)  & $0.591$ & $0.759$ & $0.002$ & $0.206$ & $-$     \\ \hline
\end{tabular}
}
\label{table:log_ratio}
\end{table}

\begin{figure}[ht!]
\centering
\includegraphics[width=\columnwidth,clip=true]{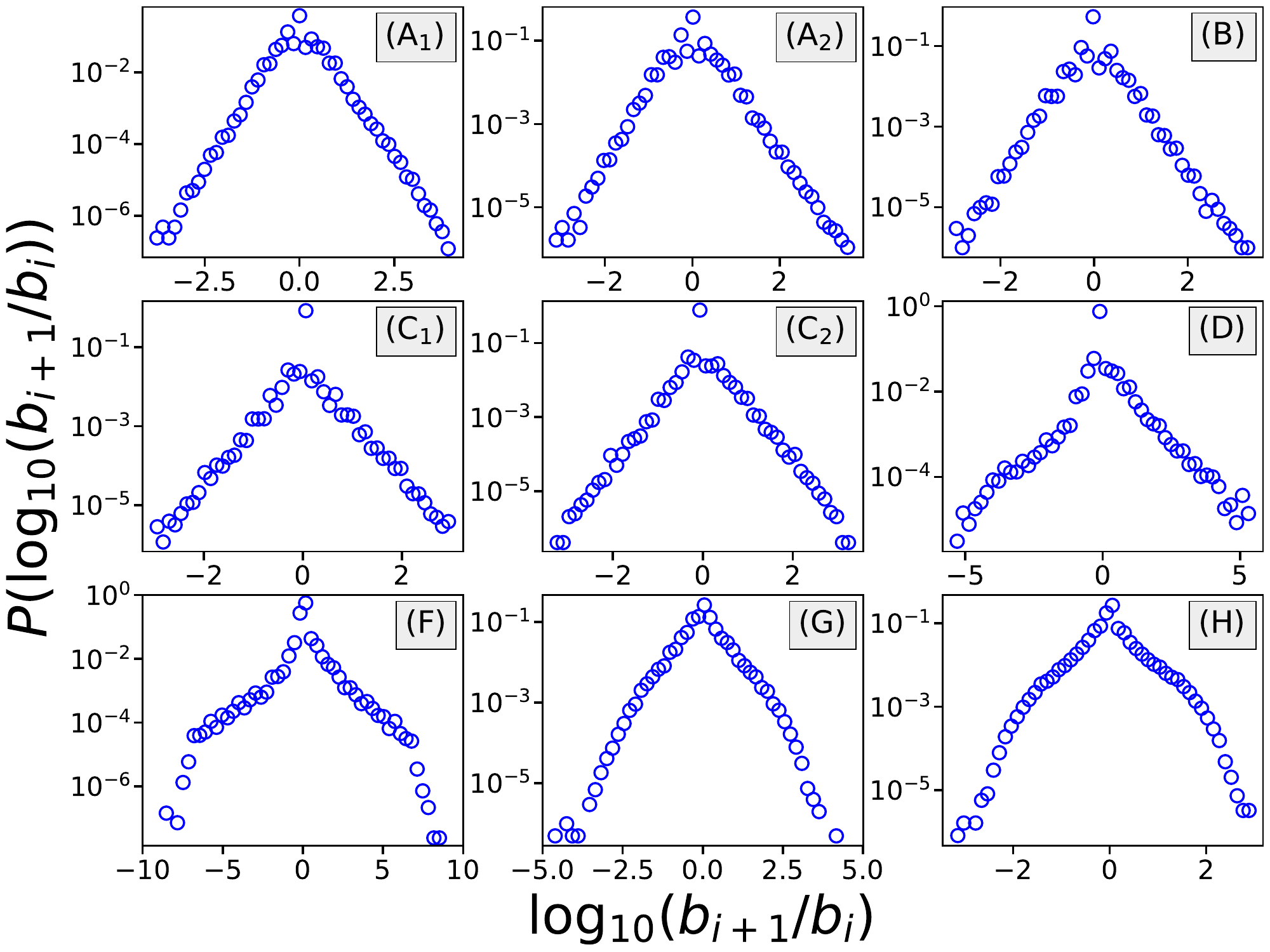}
\caption{
\label{fig:log_ratio}
The distribution of the logarithmic of the ratio (log-ratio) between
consecutive bet values. For games (A, B, C), the log-ratio can be described
by a Laplace distribution. For games (D, F, G, H), the log-ratio presents
bell-shaped distribution. In general, the distributions are symmetric with
respect to the y-axis, except in games (D) and (F). The $x$-coordinate $\log_{10}(b_{i+1}/b_i)$ is proportional to the parameter $\nu$.
}
\end{figure}

Further analysis of the distribution of $\nu$ shows an
exponential decay on both of its tails, see Fig.~\ref{fig:log_ratio}.
This means that $\nu$ approximately follows a Laplace distribution.
However, compared to a Laplace distribution,
the empirical log-ratio distribution shows a much higher probability
at $\nu=0$, whose value can be found in the last column of
Table~\ref{table:log_ratio}.
We also observe that $\nu$ presents higher probability
densities around small integers/half-integers
and their inverses. Due to the existence of these differences,
we will skip the parameter fitting for the distribution of $\nu$.
The high probability of staying on the same wager indicates that  betting with fixed wager is one of the
common strategies adopted by gamblers.

Meanwhile, the high positive auto-correlations, along with the higher
probability densities at small integers/half-integers and their inverses,
provide evidence that gamblers often follow a multiplicative
process when wagering. The multiplication process can be
explained by the wide adoption of multiplicative betting systems.
``Betting system'' here refers to the strategy of wagering where the
next bet value depends on both the previous bet value and the previous
outcome~\cite{Dubins2014, Epstein2012}. Although betting systems will not
provide a long-term benefit, as the expected payoff will always
be $0$ in a fair game, still they are widely adopted among gamblers.
A well-known multiplicative betting system is the Martingale
(sometimes called geometric progression)~\cite{Epstein2012}.
In Martingale betting, starting with an initial wager,
the gambler will double their wager each time they lose one round,
and return to the initial wager once they win.
Martingale is a negative-progression betting system where the gambler will increase their wager when they lose and/or decrease
their wager when they win.

Apart from multiplicative betting, there are many other types of betting
systems, such as additive betting and linear betting~\cite{Epstein2012}.
The reasons why multiplicative betting systems are
dominant in our datasets are: 1) Martingale is a well-known betting
system among gamblers; 2) Many online gambling websites provide a
service for changing the bet value in a multiplicative way. For example,
for the Crash games csgofast-Crash (C) and ethCrash (D), both websites
provide a simple program for automatically wagering in a multiplicative
way. For the Roulette games and Coinroll (F), the websites provide an
interface with which the gambler can quickly double or half their wager.
However, for Satoshi Dice (E) and csgospeed-Jackpot (G), no such function
is provided, yet we still observe similar results, indicating that
gamblers will follow a multiplicative betting themselves.

Fig.~\ref{fig:log_ratio} provides us with the distribution of $\nu$, however, it
will not tell us whether the gamblers adopt the negative/positive-progression
betting systems. Therefore we further analyze the effect on the bet values
of winning/losing a round.
How the gamblers adjust their wager after winning/losing rounds is shown in
Table~\ref{table:win-lose}. We can see that although there is a high
probability for sticking to the same bet values, the most likely outcome
after losing a round is that the gambler increases their wager.
When winning one round, gamblers are more likely to decrease their wager.
This means that negative-progression strategies are
more common among gamblers than positive-progression strategies.

\begin{table}[ht!]
\centering
\caption{Statistics about how gamblers change their bet values after winning/losing rounds.
Apart from fixed-wagering betting, a comparison between the probabilities suggests gamblers
prefer negative-progression betting rather than positive-progression betting.
See the caption of Table~\ref{table:log_ratio} for some additional details.}
\resizebox{\textwidth}{!}{%
\begin{tabular}{|c||c|c|c||c|c|c|}
\hline
\multirow{2}{*}{Dataset} & \multicolumn{3}{c||}{After Losing} &
\multicolumn{3}{c|}{After Winning} \\ \cline{2-7}
& $P(b_{i+1}>b_i)$ & $P(b_{i+1}=b_i)$ & $P(b_{i+1}<b_i)$
& $P(b_{i+1}>b_i)$ & $P(b_{i+1}=b_i)$ & $P(b_{i+1}<b_i)$ \\ \hline
csgofast-Double (A)      & $0.432$ & $0.319$ & $0.249$ & $0.228$ & $0.383$ & $0.388$ \\ \hline
csgofast-X50 (B)         & $0.293$ & $0.500$ & $0.207$ & $0.167$ & $0.541$ & $0.292$ \\ \hline
csgofast-Crash (C)       & $0.201$ & $0.685$ & $0.114$ & $0.076$ & $0.854$ & $0.069$ \\ \hline
ethCrash (D)             & $0.566$ & $0.401$ & $0.033$ & $0.079$ & $0.690$ & $0.231$ \\ \hline
Coinroll (F)             & $0.560$ & $0.377$ & $0.061$ & $0.121$ & $0.606$ & $0.274$ \\ \hline
csgospeed-Jackpot (G)    & $0.478$ & $0.159$ & $0.374$ & $0.415$ & $0.104$ & $0.480$ \\ \hline
\end{tabular}%
}
\label{table:win-lose}
\end{table}

\subsection*{Risk attitude}
We now turn to the following question:
When a player is allowed to choose the odds themselves in a near-fair game,
how would they balance the risk and potential return? Higher odds means
a lower chance of winning and higher potential return, for example,
setting odds of $10$ means that the winning chance is only $1/10$,
but the potential winning payoff equals $9$ times the original wager.
In our analysis, we can examine such behaviors based on the gambling logs
from Crash and Satoshi Dice games.
For the Crash game only CSGOFAST.COM provides the player-selected odds
even when players lose that round, whereas for the Satoshi Dice game
only Coinroll accepts player-selected odds.
We will therefore focus on the data collected on these two websites. For the
Crash game on CSGOFAST.COM, the odds can only be set as multiples of $0.01$,
whereas for the Satoshi Dice game on Coinroll the odds can be set to
$0.99 \cdot 65536/i$ where $i$ is a positive integer less than $64000$.
To simplify our modeling work, we will convert the odds on Coinroll
to be multiples of $0.01$ (same as for the Crash game).

\begin{figure}[ht!]
\centering
\includegraphics[width=.75\columnwidth,clip=true]{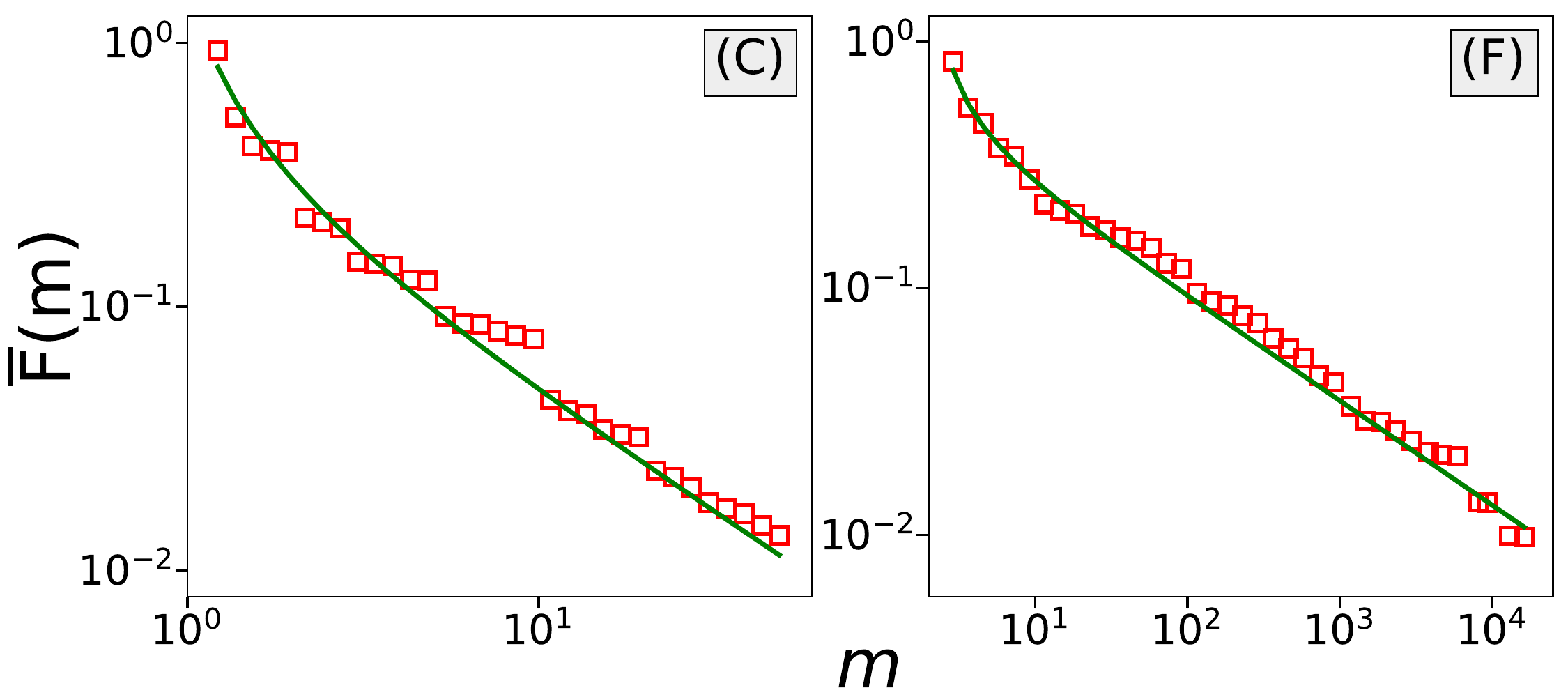}
\caption{\label{fig:odds}
Odds distributions can be well-fitted by
truncated shifted power-law distributions.
}
\end{figure}

It turns out that in both cases
the odds can be modeled with a truncated shifted power-law distribution,
\begin{equation}
\displaystyle P(m) = \left\lbrace \begin{split}
\displaystyle &\frac{\left(\ m - \delta \right)^{-\alpha}}{\zeta(\alpha, \ m_\text{min} -\delta)},
\ &\text{for}\ m_\text{min}\le m < m_\text{max}, \\
\displaystyle &\frac{\zeta(\alpha,\ m_\text{max} -\delta)}{\zeta(\alpha, \ m_\text{min} -\delta)},
\ &\text{for}\ m = m_\text{max},
\end{split}\right.
\end{equation}
where $\zeta(\cdot,\cdot)$ is the incomplete Zeta function,
and $m_\text{max}$ is the upper truncation. Note that there is a jump
at $m_\text{max}$, meaning that the players are more likely to place bets
on the maximum allowed odds than on a slightly smaller odds.
The estimated parameters $\alpha=1.881$, $\delta=0.849$, and $m_\text{min}=1.15$ for csgofast-Jackpot on CSGOFAST.COM,
whereas for Coinroll the parameters are found to be $\alpha = 1.423$, $\delta = 2.217$, and $m_\text{min} = 2.58$.
From the comparison between the CCDFs of empirical data and fitting curves,
as shown in Fig.~\ref{fig:odds}, we can see that the truncated shifted
power law can capture the overall decaying trends of odds distribution.
The stepped behavior results
from the gamblers' preference of simple numbers.

A distribution that is close to a power law indicates that
a gambler's free choice of odds displays scaling characteristics (within
the allowed range) in near-fair games. It also means that when gamblers
are free to determine the risks of their games,
although in most times they will stick to low risks,
showing a risk-aversion attitude, they still present a non-negligible
probability of accepting high risks in exchange for high potential returns.
The scaling properties of risk attitude might not be unique to gamblers,
but also may help to explain some of the risk-seeking behaviors in
stock markets or financial trading.

We now re-examine the distributions from the point of view of estimating
the crash point $m_C$ (Satoshi Dice games can be explained with the same
mechanism). The true distribution
of $m_C$ generated by the websites follow a power-law decay with an
exponent of $2$ (with some small deviation due to the house edge).
Meanwhile, a closer look at the fitted exponents listed above gives us two empirical exponents
of $1.423$ and $1.881$, both of which are smaller than $2$.
The smaller exponents reveal that gamblers believe that they have a
larger chance to win a high-odds game than they actually do.
Or equivalently, it means the gamblers over-weight the winning chance of
low-probability games.
At the same time, the ``shifted'' characteristics here lead to more bets
on small odds, which also indicates that the gamblers over-estimate the
winning chance of high-probability games. As a result,
they under-weight the winning chances of mild-probability games.
These are clear empirical evidence of probability weighting among
gamblers, which is believed to be one of the fundamental mechanisms in
economics~\cite{Barberis2012}.

\subsection*{Wealth distribution}
\begin{figure}[ht!]
\centering
\includegraphics[width=.5\columnwidth,clip=true]{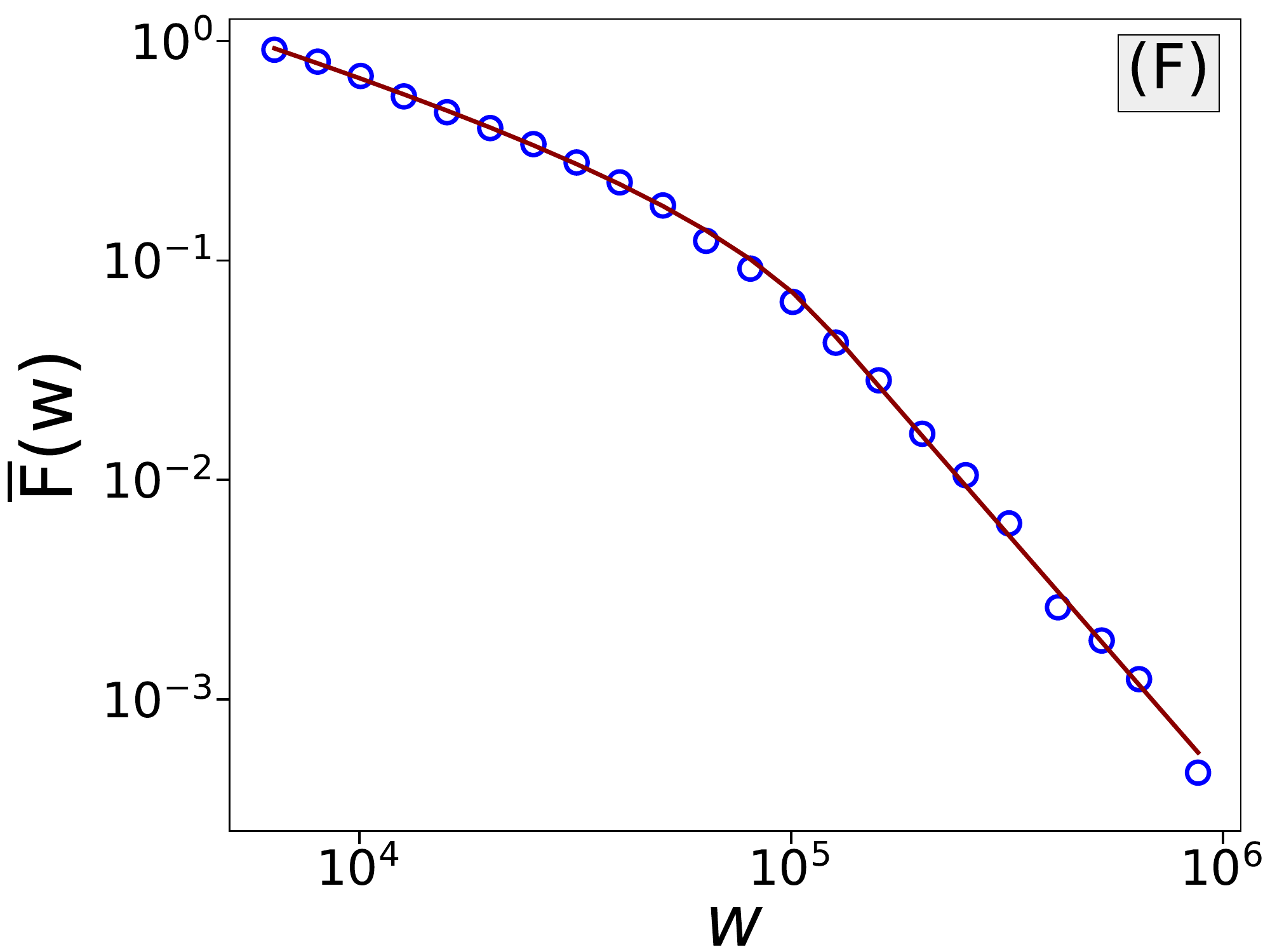}
\caption{
\label{fig:wealth}
The tail of the wealth distribution of Bitcoin gamblers follows a
pairwise power-law distribution.
}
\end{figure}

In the previous study of skin gambling~\cite{Wang2018}, we pointed out that
the wealth distribution of skin gamblers shows a pairwise power-law tail.
This time, by considering the players' deposits to their wallets on a gambling site as
the wealth data, we find that the pairwise power-law tails are
also observed for bitcoin gambling.
We find that on the gambling website Coinroll, starting from $5660$ cents,
the players' wealth distribution follows a pairwise power-law distribution,
with the power of the first regime to be $1.585$,  and the power of the
second regime to be $3.258$, see Fig.~\ref{fig:wealth}.
The crossover happens at $1.221\times 10^5$ cents.
As both wealth distributions of skin gambling and bitcoin gambling can be
approximated by a pairwise power distribution, we believe that it is a good
option for modeling the tails of gambler wealth distribution in different
scenarios.

\subsection*{Removing effects due to inequality in the number of bets}
\begin{figure}[ht!]
\centering
\includegraphics[width=\columnwidth,clip=true]{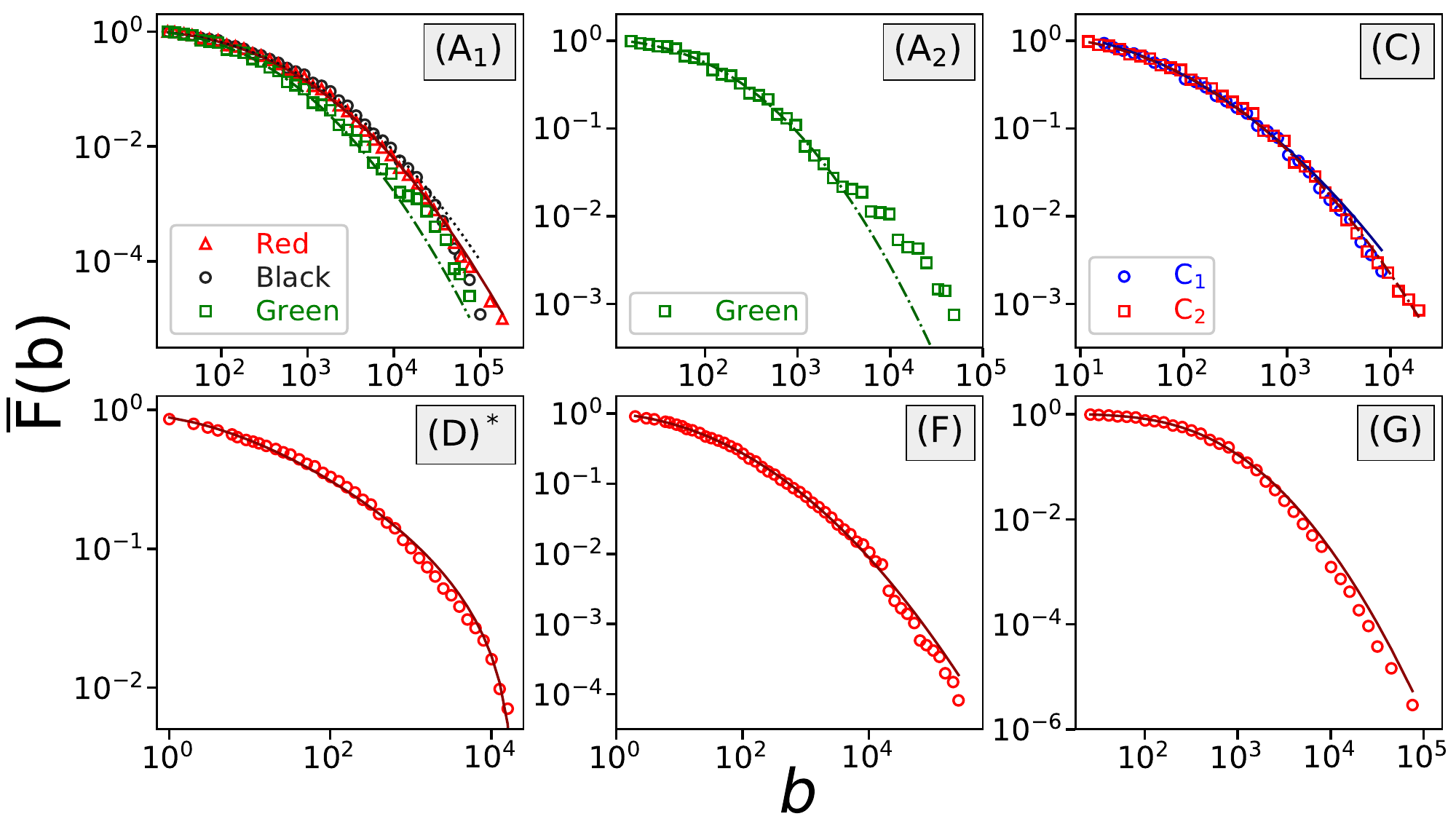}
\caption{\label{fig:wager_distr_top}
The wagers obtained from random sampling of top gamblers' bets still
present log-normal distributions, although there are some observable
deviations.
}
\end{figure}

In the above sections, we have analyzed the distributions of several quantities
at the population level. However, 
there is a huge inequality of the number of placed bets among gamblers.
We therefore wonder whether those distributions we obtain result from
the inequality of number of bets among individuals.
To remove the effects of this inequality, we randomly sample in each dataset
the same number of bets from heavy gamblers.
We re-analyze the wager distribution and odds distribution with the sample
data to see if we obtain the same distribution as before.
In each dataset we randomly sample $500$ bets from
each of those gamblers who placed at least $500$ bets above $b_\text{min}$
given in Table~\ref{table:wager-distribution}. Some datasets are excluded
here as either they do not have enough data or we cannot identify individual
gamblers. When re-analyzing the odds distribution, to ensure we have enough data,
we respectively sample $100$ and $2000$ bets from each of those gamblers in
games (C) and (F) who have at least $100$ and $2000$ valid player-selected odds
above $m_\text{min}$.
According to the results in Fig.~\ref{fig:wager_distr_top},
after removing the inequality the wager distributions can still be
approximated by log-normal distributions, but some deviation can be observed.
Similarly, the odds distributions 
again follow truncated shifted power-law distributions after removing the inequality.
These results demonstrate that the shape of the distributions we obtained
in the above sections is not a result of the inequality of the number of bets.

Now our question becomes whether the conclusion regarding the distribution
at the population level can be extended to the individual level. Here due to
the limitation of data, we will only discuss the wager distribution.
Analyzing the individual distribution of top gamblers, we find that
although heavy-tailed properties can be widely observed at the individual
level, only a small proportion of top gamblers presents log-normal distributed
wagers. Other distributions encountered include log-normal distributions, power-law distributions,
power-law distributions with exponential cutoff, pair-wise power-law
distributions, irregular heavy-tailed distributions, as well as distributions
that only have a few values.
The diversity of the wager distributions at
the individual level suggests a diversity of individual betting strategies.
Also, it indicates that a gambler may not stick to only one betting strategy.
It follows that the log-normal wager distribution
observed at the population level is very likely an aggregate result.

\subsection*{Diffusive process}
For an individual player's gambling sequence
we define ``time'' $t$ as the number of bets one
player has placed so far, and define as net income the sum of the
payoffs of those bets.
In all the games we analyze, there are
only two possible outcomes: a win or a loss.
The player's net income will change each time they place a bet in
a round, with the step length to be the payoff from that bet.
We can treat the change of one player's net income as a random walk
in a one-dimensional space 
(see Fig. 1 in Ref. \cite{Wang2018} for an example of such a trajectory).
The time $t$ will increase by $1$ when the player places a new bet,
therefore the process is a discrete-time random walk.

\begin{figure}[ht!]
\centering
\includegraphics[width=\columnwidth,clip=true]{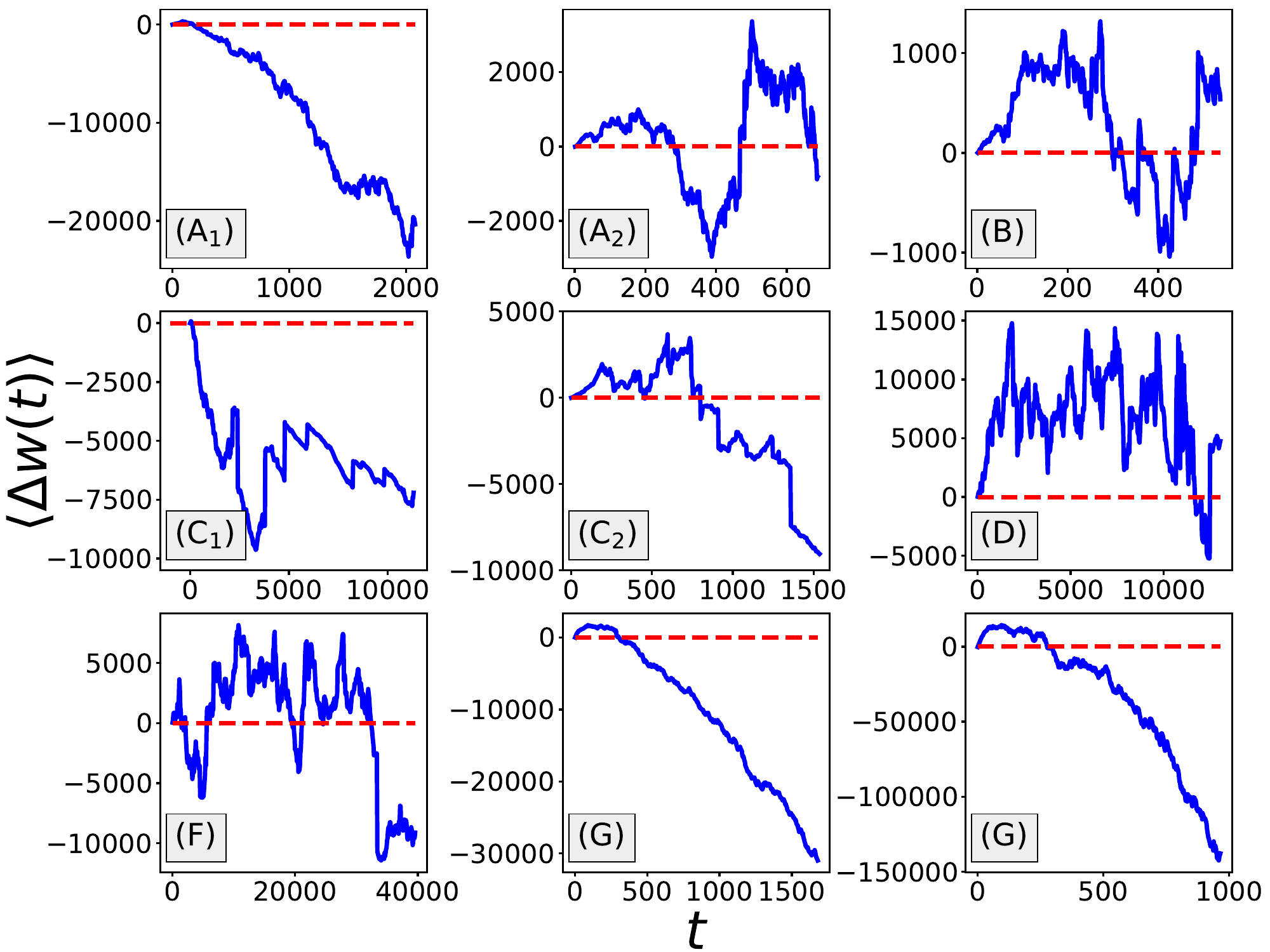}
\caption{
\label{fig:netincome}
Change of the mean net income with time for the different datasets. Most of the datasets present a decreasing net
income as time $t$ increases. Each point is obtained from an average over at least $200$ players.}
\end{figure}

Now, let us focus on the analysis of the diffusive process of the
gamblers' net incomes, starting with the analysis of the
change of the mean net income  with the number of rounds played (time),
$\left\langle \Delta w(t) \right\rangle = \left\langle w(t) - w_0 \right\rangle =
\left\langle \sum\limits_{i=1}^t o_p(i) \right\rangle,$
where $w_0$ is the player's initial wealth, $w(t)$ is the player's
wealth after attending $t$ rounds, and $o_p(i)$ is the payoff
from the $i_{th}$ round the player attended. $\langle \cdot \rangle$
represents an ensemble average over a population of
players placing bets. In the rest of this paper,
$\langle \cdot \rangle$ will always be used for representing an ensemble average.
In Fig.~\ref{fig:netincome} we show the change of $\left\langle \Delta w(t) \right\rangle$
over time. In most of the datasets, players' mean net income decreases over time,
which suggests that in general players will lose more as they gamble more.
At the same time, in some datasets such as Ethcrash (D) and Coinroll (F), large fluctuations
can be observed. 

An useful tool for studying the diffusive process is the
ensemble-averaged mean-squared displacement (MSD), defined as
\begin{equation}
\left\langle\Delta w^2(t) \right\rangle =
\left\langle\left(w(t)-w_0\right)^2\right\rangle =
\left\langle \left(\sum\limits_{i=1}^t o_p(i)\right)^2 \right\rangle,
\end{equation}
For a normal diffusive process,
$\left\langle\Delta w^2(t) \right\rangle \sim t$,
otherwise an anomalous diffusive behavior prevails. More specifically, when the MSD growth is
faster (respectively, slower) than linear, superdiffusion (respectively, subdiffusion) is observed.

\begin{figure}[ht!]
\centering
\includegraphics[width=\columnwidth,clip=true]{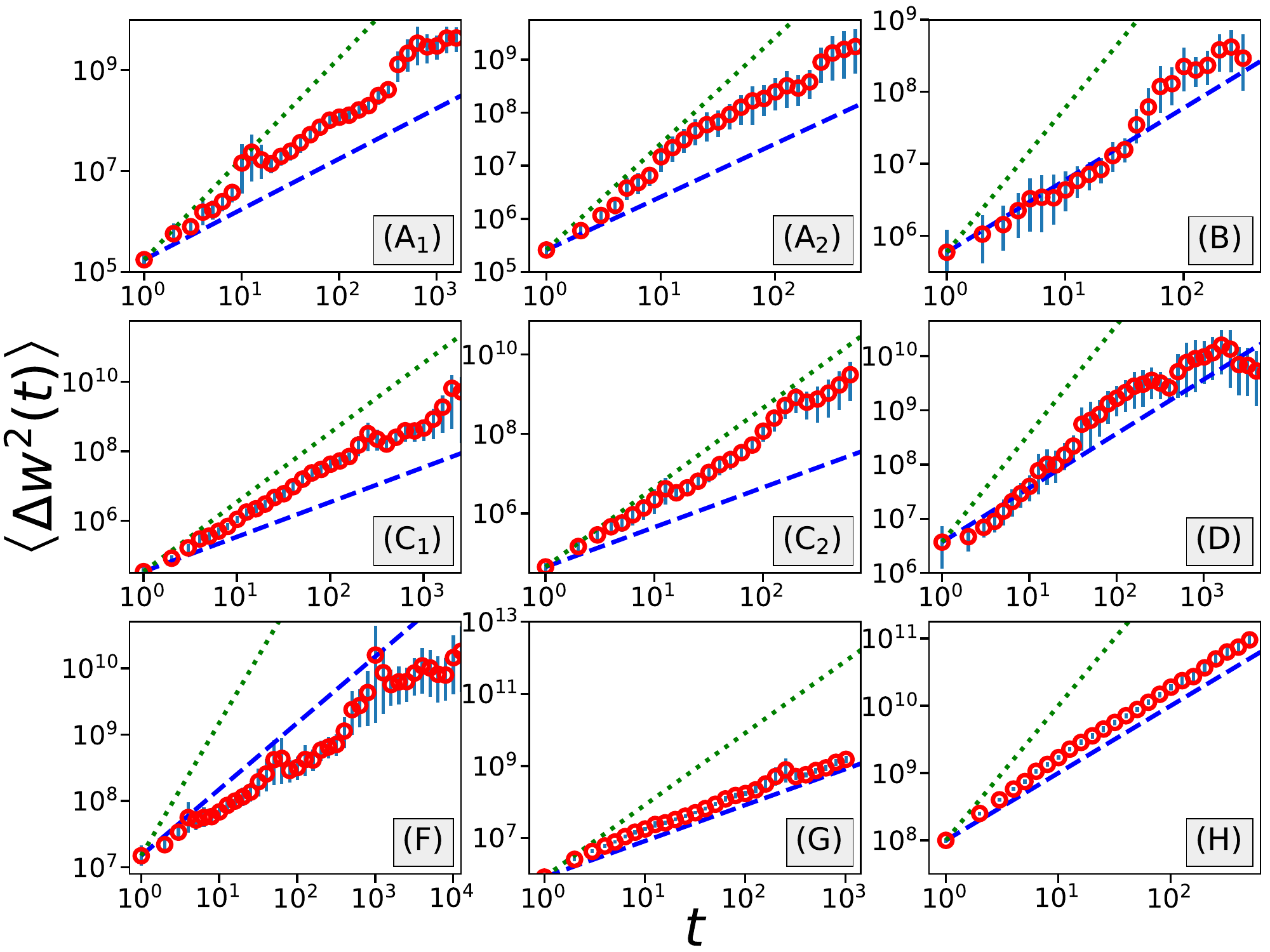}
\caption{
\label{fig:msd}
The growth of ensemble-averaged mean-squared displacement in different datasets presents
different diffusive behaviors. In the figures, the error bars represent
$95\%$ confidence intervals, blue dashed lines follow linear
functions (slope~$=1$), and green dotted lines follow quadratic functions
(slope~$=2$).}
\end{figure}

In Fig.~\ref{fig:msd}, we present the growth of the ensemble-averaged MSD
against time for each of the datasets.
To reduce the coarseness, MSD curves are smoothed with log-binning
technique. The error bars in Fig.~\ref{fig:msd} represent $95\%$
confidence intervals computed with bootstrapping using $2000$
independent re-sampling runs. It is interesting to see that
for different datasets we observe different diffusive behaviors.
For games csgofast-Crash (C) we observe that the
MSD grows faster than a linear function, suggesting
superdiffusive behavior.
Meanwhile, for games csgofast-Double (A), ethCrash (D), csgospeed (G),
and csgofast-Jackpot (H),
the MSD first presents a superdiffusive regime,
followed by a crossover to a normal diffusive regime.
For games csgofast-X50 (B) and Coinroll (F),
although the ensemble-averaged MSD roughly
presents a linear / sublinear growth, a careful inspection shows
that both curves consist of several convex-shaped regimes, indicating
a more complex behavior. Convex-shaped regimes can also be observed
in csgofast-Crash games (C).

In Ref. \cite{Wang2018} we argued that the crossover from a
superdiffusive regime to a normal diffusive regime in a parimutuel game
is due to the limitation of individuals' wealth and the conservation of
total wealth. Similar crossovers are observed in games (G) and (H),
two parimutuel betting games,
where the same explanation can be applied.
On the other hand, this crossover is also found in a Roulette game and in a Crash game,
where there is no interaction among gamblers. The limitation of an individual's wealth
can still be a partial explanation, but the conservation of total wealth no longer
holds. A different explanation needs to be proposed to model this crossover.

In the following we briefly discuss how we can obtain from 
gambling models the different diffusive processes observed in the
data.
We will not attempt to reproduce the parameters we obtained from
the gambling logs, but rather try to explore the possible reasons
for the anomalous diffusion we reported.

For a gambling process, if the gambler's behavior is independent among
different rounds, i.e., the wager and odds are respectively independent
and identically distributed (IID), with no influence from the previous
outcomes, and if the wager $b$ has finite variance and the odds $m$ has
finite mean, then MSD's growth will be a linear function of time $t$:
\begin{equation}\label{eq:linear-msd}
\displaystyle \left\langle \Delta w^2(t) \right\rangle =
\left\langle\left(w(t)-w_0\right)^2\right\rangle
 =(\langle m\rangle-1)\left\langle b^2\right\rangle t,
\end{equation}
where $\langle m\rangle$ is the mean value of odds distribution and
$\left\langle b^2\right\rangle$ is the second moment of the wager distribution.
But normal diffusion is only found in few datasets, the remaining datasets
presenting anomalous diffusion which conflicts with the IID assumption.

\begin{figure}[ht!]
\centering
\includegraphics[width=.75\columnwidth,clip=true]{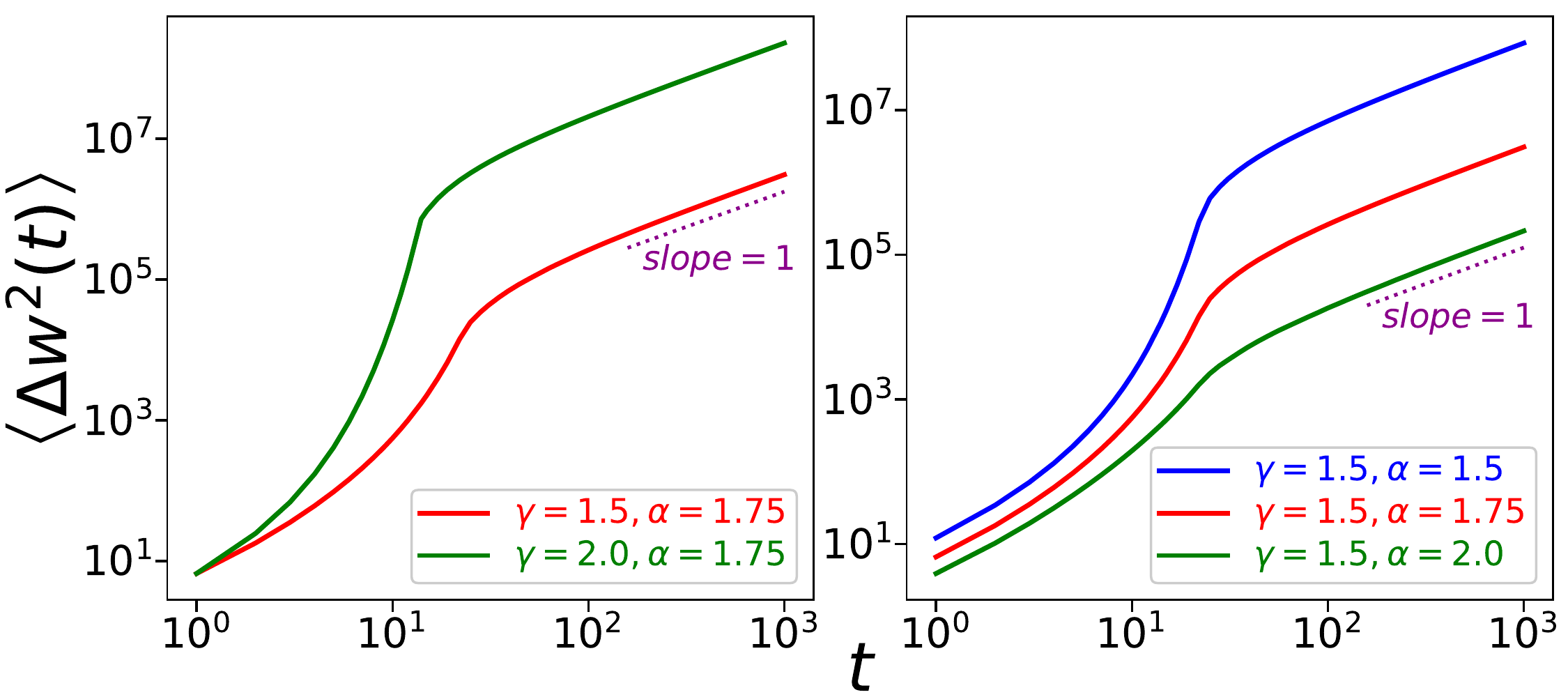}
\caption{
\label{fig:martingale}
A betting system similar to Martingale will lead to a crossover from
superdiffusion to normal diffusion according to the growth of mean-squared
displacement. Comparison between curves of different parameters shows that higher
$\gamma$ and lower $\alpha$ both will lead to a higher chance of huge losses/winnings.
}
\end{figure}

Having shown the popularity of betting systems among gamblers,
we would like to check how different betting systems affect diffusive
behaviors. First, we simulate gamblers that follow Martingale strategies in
a Crash game. We assume that the selection of odds follows a power-law
distribution with an exponent $\alpha$,
with a minimum odds of $1$ and a maximum odds of $50$, where the maximum
odds is set to ensure a finite mean of the odds distribution.
Starting from a minimum bet of $1$, we multiply wagers by a ratio
$\gamma$ each time the gamblers lose one round and return to the minimum
bet each time they win. Once the wager reaches a preset
maximum bet value $10000$, we reset the gambler with a minimum bet.
MSD obtained from 10 billion individual simulations is shown in
Fig.~\ref{fig:martingale}. Different curves correspond to different exponents
in odds distribution. We can see that the MSD initially presents
an exponential-like growth, before the growths reduce to a linear function.
It is easy to explain the exponential growth since
many gamblers lose the rounds and therefore increase their wager
by the factor $\gamma$, which leads to an increase in the average bet value.
The superdiffusion here suggests that Martingale strategy increases
gamblers' risks of huge losses.
Considering the wide adoption of Martingale among gamblers, this could
be a reason for the superdiffusion as well as the crossover to normal diffusion
we found in several datasets. Comparison of the MSD curves of different
$\alpha$ suggests that a more aggressive risk attitude leads to a higher risk
of huge losses (as well as higher potential winnings).

\begin{figure}[ht!]
\centering
\includegraphics[width=.75\columnwidth,clip=true]{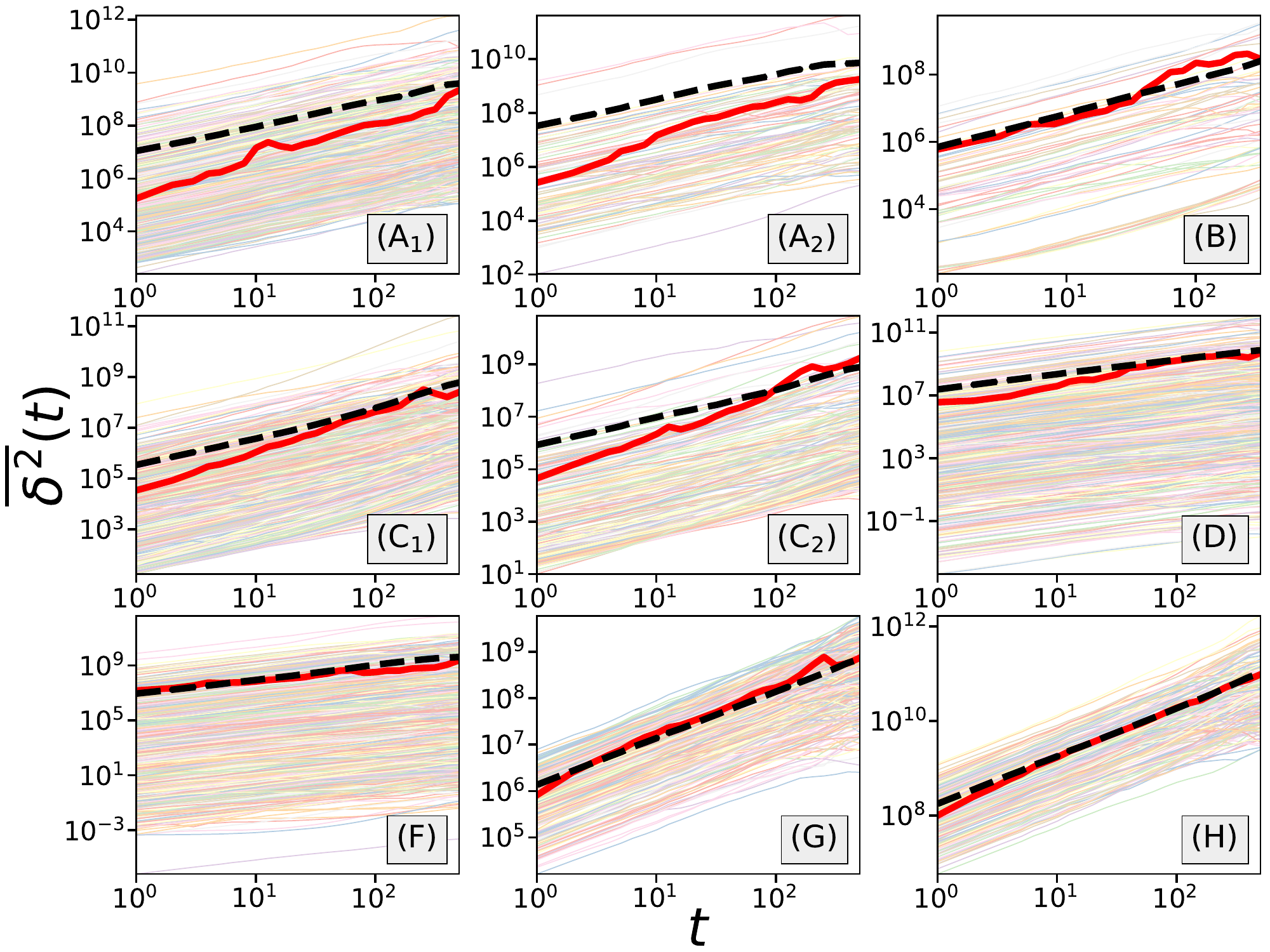}
\caption{
\label{fig:msd_t}
The growth of the time-averaged MSD for individual gamblers, presented as thin lines, suggests
diverse betting behaviors at the individual level. The comparison between
$\left\langle \overline{\delta^2}(t) \right\rangle$ (thick dashed black lines) and
$\left\langle\left( \Delta w(t) \right)^2 \right\rangle$ (thick full red lines)
reveals that these quantities are different for most games,
with the exception of the Coinroll (F), csgospeed (G) and csgofast-Jackpot (H) games.
Players who played less than $1000$ rounds are filtered out in each dataset.
}
\end{figure}

Next we examine the ergodicity of the random walk process of net income by computing the time-averaged
mean-squared displacement and the ergodicity breaking parameter. The time-averaged MSD is defined as
\begin{equation}
\displaystyle \overline{\delta^2}(t) = \frac{1}{T-t} \sum\limits_{k=1}^{T-t} \left(w(k + t) - w(k)\right)^2, 
\end{equation}
where $T$ is the length of the player's betting history, i.e. total number of rounds they attend,
and $\overline{\cdots}$ is used for representing a time average.
To calculate the time-averaged MSD, we need to make sure the player has played enough rounds so that we have
a long enough series of net income data, therefore in each dataset we filter out the players who played less than $T=1000$ rounds.
As shown in Fig.~\ref{fig:msd_t} the time-averaged MSD shows huge deviations from player to player,
suggesting diverse betting behaviors at the individual level.
At the same time, comparison between
the ensemble-averaged time-averaged MSD $\left\langle \overline{\delta^2}(t) \right\rangle$
and the ensemble-averaged MSD $\left\langle\Delta w^2(t)\right\rangle$
shows clear deviations in most datasets, except in the Coinroll (F), csgospeed (G) and csgofast-Jackpot (H) games.
To further examine breaking of ergodicity, we have calculated the ergodicity breaking parameter $\rm{EB}$~\cite{Cherstvy2013, Cherstvy2014, Cherstvy2019}
defined as
\begin{equation}
\displaystyle \rm{EB}(t) = \left\langle \left(\overline{\delta^2}(t)\right)^2\right\rangle \bigg/ \left\langle\overline{\delta^2}(t)\right\rangle^2 - 1.
\end{equation}

\begin{figure}[ht!]
\centering
\includegraphics[width=.75\columnwidth,clip=true]{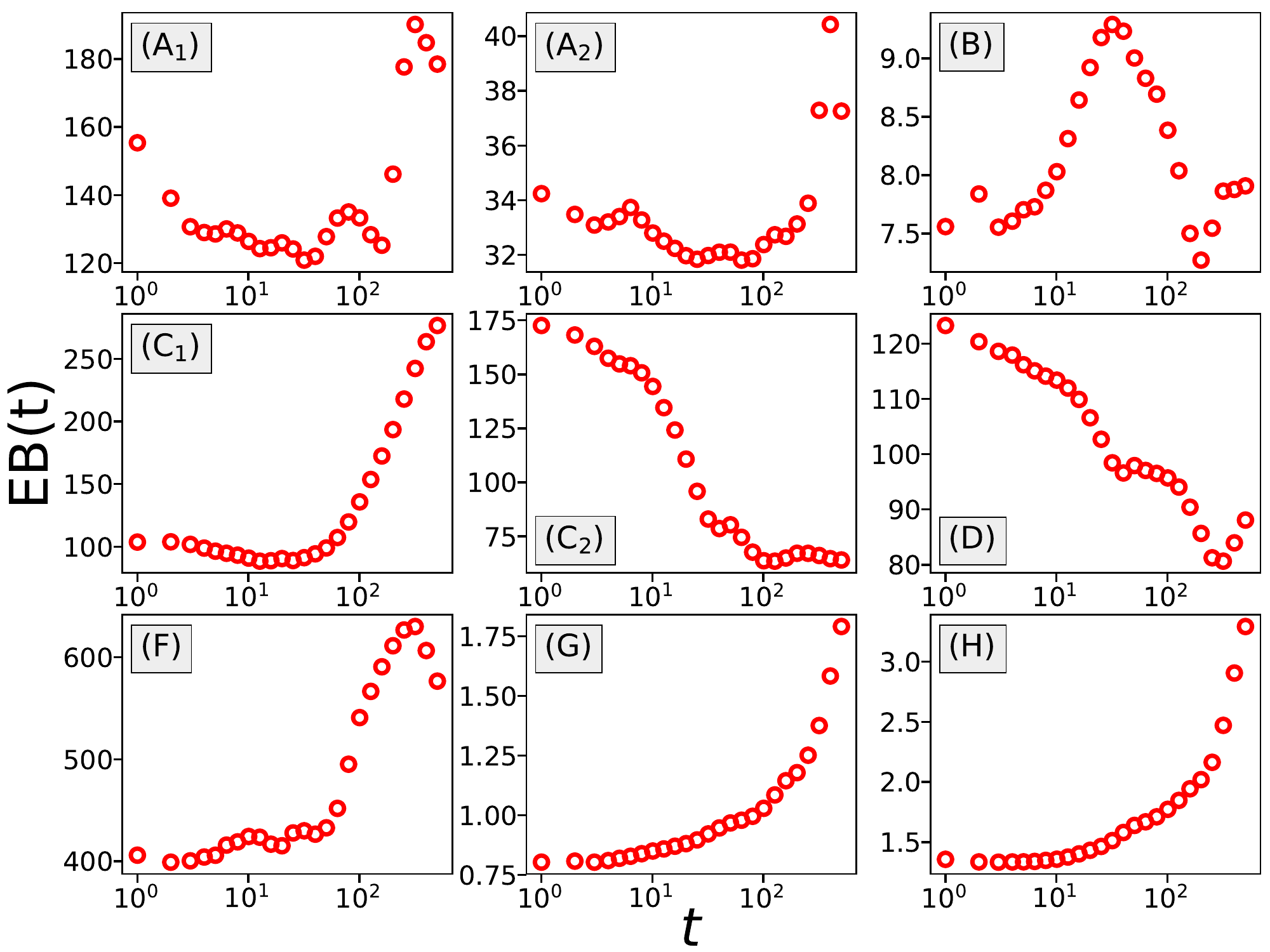}
\caption{
\label{fig:eb}
The change of the ergodicity breaking parameter with time. For all games, with the
exception of the games csgospeed (G) and csgofast-Jackpot (H),
${\rm{EB}}$ is found to be much larger than $0$, suggesting non-ergodic behavior.}
\end{figure}

For an ergodic process, the parameter $\rm{EB}$ should be close to $0$. However, as shown in Fig.~\ref{fig:eb},
in most datasets, with the exception of csgospeed (G) and csgofast-Jackpot (H), $\rm{EB}$ is large.
It follows that non-ergodicity is observed in most games
and that gambling processes indeed often deviate from normal diffusion,
which further highlights the complexity of human gambling behavior.

\begin{figure}[ht!]
\centering
\includegraphics[width=.75\columnwidth,clip=true]{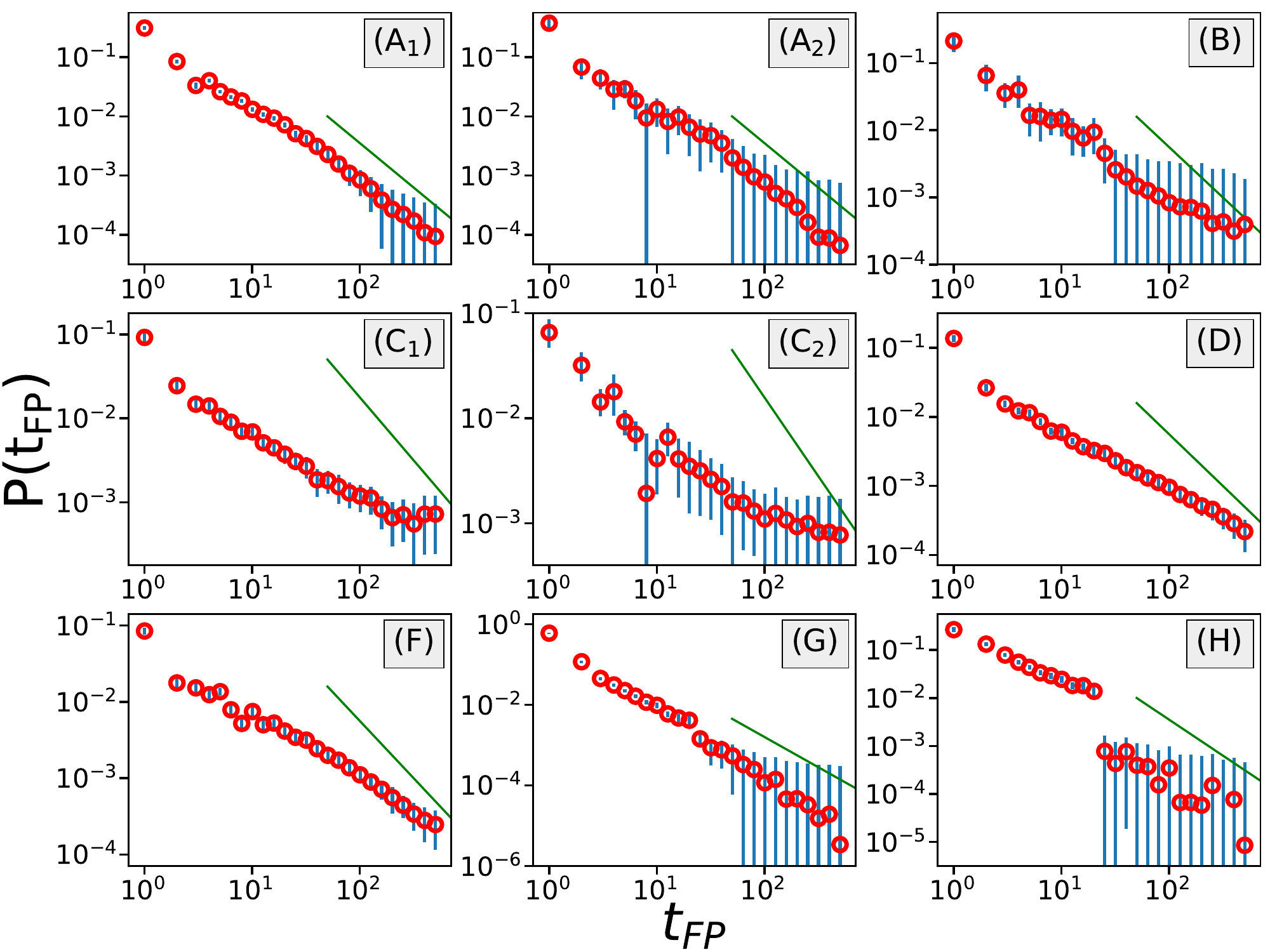}
\caption{
\label{fig:fpt}
The tails of first-passage time distributions for the
different datasets indicate different diffusive behaviors.
The green lines represent a power-law decay with an exponent $3/2$.
The blue error bars indicate $95\%$ confidence intervals.
Only gamblers who attended more than 1000 rounds of games have been included in these calculations.
}
\end{figure}

Another way to examine the diffusive behavior of a process is through the analysis of the first-passage time distribution. The first-passage time $t_{FP}$ is the time required for a random walker at location $w$ to leave the region $[w-V_{FP},w +V_{FP}]$ for the first time, where $V_{FP}$ is the target value or first-passage value. The first-passage time distribution $P(t_{FP})$~\cite{Inoue2007
, Wang2018}, defined as the survival probability that the random walker, who is located at $w$ at time $t_0$, stays within range $[w-V_{FP},w +V_{FP}]$ up to time $t = t_0 + t_{FP}$, can be calculated from the expression
\begin{equation}
P(t_{FP}) = \lim\limits_{T \longrightarrow \infty} \frac{1}{T} \sum\limits_{k=1}^T \Theta \left( \left| 
w(k+t_{FP}) - w(k) \right| - V_{FP} \right) - \lim\limits_{T \longrightarrow \infty} \frac{1}{T} \sum\limits_{k=1}^T \Theta \left( \left|
w(k+t_{FP}-1) - w(k) \right| - V_{FP} \right)~,
\end{equation}
where $\Theta(\cdot)$ is the Heaviside step function. We use $V_{FP} = 200$ (US cents),
with the exception of csgofast-Jackpot (H) for which $V_{FP}$ is chosen to be 5000.
For a normal diffusive process, the tail of $P(t_{FP})$ 
should decay with an exponent of $3/2$. In Fig.~\ref{fig:fpt} we plot the first-passage time distribution for 
each dataset, where again diverse diffusive behaviors are observed. In the games csgofast-Double (A) and 
csgofast-Jackpot (H), the tails of $P(t_{FP})$ approximately decay with an exponent of $3/2$ (see the thin 
green lines), indicating normal diffusive processes. For the game csgospeed (G), the exponent is found
to be larger than $3/2$, indicating a superdiffusive process. And in games csgofast-X50 (B), csgofast-Crash (C), 
ethCrash (D), and Coinroll (F), the exponents are clearly smaller than $3/2$, indicating a subdiffusive behavior. 
We note that
the results obtained from ensemble-averaged MSD sometimes differ from the results obtained from the
first-passage time distributions. Nonetheless, anomalous diffusive behavior is widely observed.

\begin{figure}[ht!]
\centering
\includegraphics[width=.75\columnwidth,clip=true]{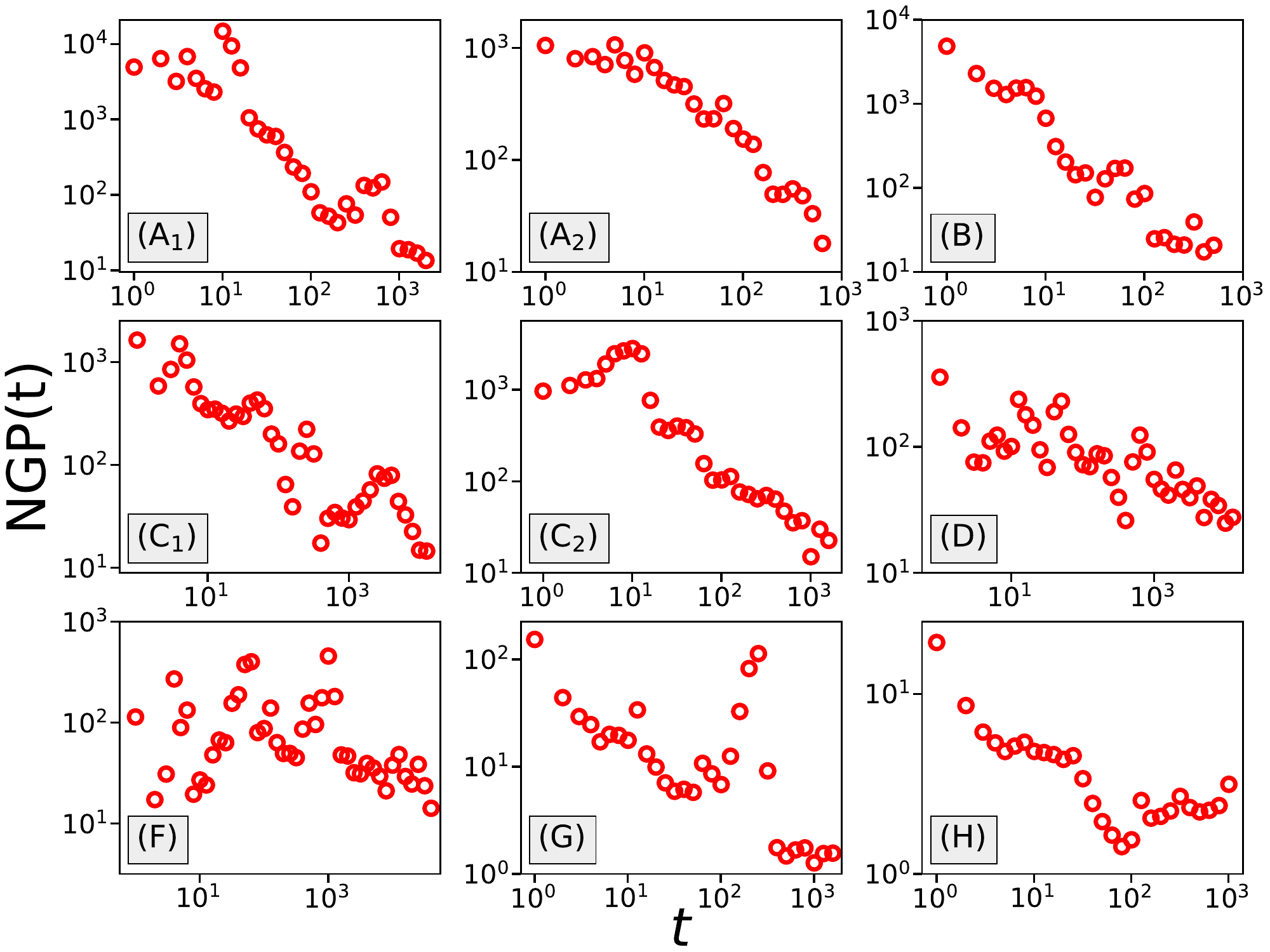}
\caption{
\label{fig:ngp}
In most datasets, except Coinroll (F), the non-Gaussian parameter shows a decreasing trend as $t$ increases. 
However, in none of the studied cases does the non-Gaussian parameter fall below the value 1.
}
\end{figure}

To confirm our conclusion about the wide existence of anomalous diffusive behavior in gambling activities, we further 
calculate the non-Gaussian parameter (NGP) \cite{Rahman1964,Toyota2011,Cherstvy2019}  
\begin{equation}
\displaystyle NGP(t) = \frac{\left\langle \Delta w^4(t)\right\rangle}{3\left\langle \Delta w^2(t)\right\rangle^2} - 1~.
\end{equation}
For a Gaussian process, the NGP should approach $0$ when $t$ gets large. In Fig.~\ref{fig:ngp} we show the  
NGP as a function of time. In most of the games, except Coinroll (F), NGP shows a clear decreasing trend as 
$t$ increases. In the game Coinroll (F), a decrease is not apparent, and most likely this game does not follow 
a Gaussian process. In the other games, although the NGP is still decreasing, we can not discriminate whether 
for large $t$ this quantity will tend to $0$ or instead reach a plateau value larger than zero.
For example, for the game csgospeed (G) the NGP seems to reach a plateau $NGP(t)\approx 1.5$ instead of 
continuing to decrease, but this could also be the consequence of insufficient data. Still, our analysis 
does not provide clear evidence for the presence of Gaussianity in gambling behaviors.

To sum up our analysis of the players' net incomes viewed as random walks,
the diverse diffusive behaviors found in the datasets
indicate that human gambling behavior is more complex than random
betting and simple betting systems. Further studies
are required in order to fully understand the observed differences.
At the individual level, as has been pointed out by Meng~\cite{Meng2018},
gamblers show a huge diversity of betting strategies,
and even individual gamblers constantly change their betting strategy.
Differences in the fractions of gamblers playing specific
betting strategies could be a reason why we see a variety of
diffusive behaviors in the datasets.

\section*{Discussion}
The quick development of the video gaming industry has also resulted
in an explosive growth of other online entertainment. This is especially true
for online gambling that has evolved quickly into a booming industry with multi-billion levels. Every day million of bets are placed
on websites all around the globe as many different gambling games are available online for gamblers.

Analysing different types of gambling games (ranging from Roulette to Jackpot games),
we have shown that log-normal distributions can be widely used
to describe the wager distributions of online gamblers at the aggregate level.
The risk attitude of online gamblers shows scaling properties too,
which indicates that although most gamblers are risk-averse, they sometime
will take large risks in exchange for high potential gains.

Viewing the gamblers' net income as a random walk in time (where for each gambler time is increased by one 
unit every time they play a game), we can analyze the mean-squared displacement of net income and related quantities
like the ergodicity breaking parameter or the non-Gaussian parameter
with the goal to gain an understanding of the gamblers' betting strategies through the diffusive behaviors emerging 
from the datasets. For some games the mean-squared displacement and the first-passage time distribution
reveal a transition from superdiffusion to normal diffusion as time increases. For all games the ergodicity breaking parameter
and the non-Gaussian parameter reveal deviations from normal diffusion. All this indicates that
gamblers' behaviors are very diverse and more complex than what would be expected from simple
betting systems. We speculate that one of the reasons for the observed diverse diffusive behaviors at the aggregate level
can be found in the differences in the fractions of gamblers playing specific betting strategies,
but more work is required to fully
understand the gamblers' complex behaviors.

\section*{Methods}

\subsection*{Detailed rules of the different games}
\subsubsection*{Roulette}
We focus on a simplified version of Roulette games
that appears in online casinos,
where a wheel with multiple slots painted with different colors will be spun,
after which a winning slot will be selected.
The Roulette table of a traditional Roulette game is composed of
$38$ slots, among which $18$ slots
are painted in black, $18$ slots are painted in red, and two slots (``0'' and ``00'')
are painted in green. The online Roulette games are similar 
to the traditional ones, except that the number of colors and the number of slots
for each color might be different.
Each slot has the same probability to be chosen as the winning slot.
Players will guess the color of the winning slot before the game starts.
The players have a certain time for wagering, after which the game ends and
a winning slot is selected by the website. Those players
who successfully wagered on the correct color win, the others lose.
As the chance of winning and odds for each color are directly
provided by the website,
roulette is a fixed-odds betting game.

\subsubsection*{Crash}
``Crash'' describes a type of gambling games mainly hosted in online casinos.
Before the game starts, the site will generate a crash point $m_C$,
which is initially hidden to the players.
With a lower boundary of $1$,
the crash point is distributed approximately in an inverse square law.
The players need to place their wager in order to enter one round.
After the game starts, on the player's user interface a number,
called multiplier, will show up and gradually increase from $1$
to the predetermined crash point $m_C$, after which the game ends.
During this process, if the player ``cash-outs'' at a certain multiplier $m$,
before the game ends, they win the round; otherwise they lose.
This multiplier $m$ they cashed out at is the odds, which means when winning,
the player will receive a prize that equals his wager multiplied by $m$.
When $m_C$ is generated with a strict inverse-square-law distribution, the
winning chance exactly equals the inverse of the player-selected odds $m$.
The player can also set up the cash-out multipliers automatically
before the game starts,
to avoid the possible time delay of manual cash-out.
Since in a manual cash-out scenario, after the game starts,
the multiplier will show up on the screen,
at a given moment
the decision of the cash-out multiplier 
is based on the player's satisfaction with the current multiplier,
and involves more complicated dynamics of decision-making processes.
Meanwhile, in an auto cash-out scenario, the multiplier $m$ is chosen
before the game starts, which means the decision making is more ``static.''
Crash is also a fixed-odds betting game where the odds are player-selected.

\subsubsection*{Satoshi Dice}
Satoshi Dice is one of the most popular games in crptocurrency gambling.
In 2013, the transactions resulting from playing Satoshi Dice games
accounted for about $60\%$ of overall Bitcoin transactions~\cite{Meiklejohn2013}.
When playing Satoshi Dice, the player needs to pick a number $A$ within a range
$(0, U)$ provided by the website.
The odds can be calculated with the expression $m=U/A$.
Once the player finishes wagering, the website will pick another number $B$
which is uniformly distributed on $(0, U)$.
If $B$ is less than $A$, then the player wins the round, otherwise they lose.
Satoshi Dice is a fixed-odds betting game.
In some online casinos, players cannot choose $A$ arbitrarily, but instead,
they have to select $A$ from a preset list provided by the gambling website.
Since the odds $m$ is determined from $A$, we are more interested in
the case where the players can choose $A$ arbitrarily,
from which we can obtain a more detailed distribution of the odds $m$, which
helps us to understand the players' risk attitude.
According to the rules of Satoshi Dice games, the maximum allowed bet is
proportional to the inverse of $A$, which means the accepted range of wager
is directly related to the odds.

\subsubsection*{Jackpot}
Unlike the games discussed above, Jackpot is a parimutuel betting game,
where players gamble against each other. During the game,
each player attending the same round will deposit their wager to a pool.
The game-ending condition varies across different websites,
it could be a certain pool size, a certain amount of players,
or a preset time span.
When the game ending condition is reached, each player's winning chance
will be determined by the fraction of their wager in the wager pool,
based on which one player will be chosen as the winner by the website.
The winner will obtain the whole wager pool as the prize,
after excluding the site cut.
The odds can be calculated by the pool size divided by the player's wager,
but it is unknown to the players at the moment they wager.
In the previous study~\cite{Wang2018},
we have already discussed the player's behavior in Jackpot games of skin gambling
where in-game skins are directly used as wagers.
In this paper, we extend the analysis to a case where wagers can be arbitrary amounts of virtual skin tickets (players need to first exchange in-game skins into virtual skin tickets).

\subsection*{Data summary}
For each type of game, we collect two datasets.
In total, we analyze 8 datasets collected from
4 different online gambling websites, and the number of bet logs contained
in each dataset ranges from $0.3$ million to $19.2$ million.
Due to the high variation of market prices of
crypto-currencies and in-game skins, the wager and deposits are first converted
into US cents based on their daily market prices.

\subsubsection*{CSGOFAST}
From the skin gambling website CSGOFAST~\cite{csgofast} we collected
four datasets on the Roulette, Crash and Jackpot games
({\it csgofast-Double}, {\it csgofast-X50}, {\it csgofast-Crash},
{\it csgofast-Jackpot}) it provides.

{\it csgofast-Double} (A) is a Roulette game in which players can bet on 3
different colors (Red, Black, Green), which respectively provide odds of (2, 2, 14).
The data were collected in two different time periods,
and the only difference between them is a change of the maximum allowed bet values.
{\it csgofast-X50} (B) is also a Roulette game in which players can bet on 4 different
colors (Blue, Red, Green, Gold) with odds (2, 3, 5, 50).

{\it csgofast-Crash} (C) is a Crash game.
As we mentioned earlier, when analyzing the risk attitude of gamblers in Crash game,
we are more interested in how players set up the
odds (multiplier) with the automatically cash-out option. On CSGOFAST,
under the automatically cash-out option, players can only setup odds ranging
from $1.10$ to $50$. The interesting point about this dataset is that even
if the player loses the round, if they used the automatically cash-out option,
it still displays the player-selected odds (which is set before the game starts);
meanwhile if they used the manually cash-out option, no odds is displayed.
Therefore in early-crashed games ($m_C < 1.10$), all the displayed odds
that are larger than $1.10$ were placed with automatically cash-out option.
These displayed odds will be used in odds distribution analysis.
The data are also collected in two different periods, where the only difference is
still a change of the maximum allowed bet value.
Roulette and Crash games on CSGOFAST all use virtual skin tickets for wagering.

{\it csgofast-Jackpot} (H) is a Jackpot game, where in-game skins are directly placed as
wagers. Each skin has a market value that ranges from $3$ to $180 000$ US cents.
A player can place at most $10$ skins in one round.

\subsubsection*{CSGOSpeed}
From the skin gambling website CSGOSpeed~\cite{csgospeed}
we collected one dataset from its Jackpot game
{\it csgospeed-Jackpot} (G), in which arbitrary amounts of virtual skin tickets can
be used as wagers. The difference between datasets (H) and (G) focuses on whether the wagers
are in-game skins or virtual skin tickets.

\subsubsection*{ethCrash}
ethCrash~\cite{ethcrash} is a cryto-currency gambling website providing a Crash game {\it ethCrash} (D).
Players need to place wagers in Ethereum (ETH), one type of crypto-currency.

\subsubsection*{SatoshiDice}
SatoshiDice~\cite{satoshidice} is a cryto-currency gambling website which accepts Bitcoin Cash (BCH)
as wagers. It provides a Satoshi Dice game  {\it satoshidice} (E),
where only 11 preset odds can be wagered on, ranging from $1.05$ to $1013.74$.
Among the preset odds, we find that more than $30\%$ of the bets are placed
under the odds $1.98$, and we will analyze those bets for wager distribution.

\subsubsection*{Coinroll}
Coinroll~\cite{coinroll} is a cryto-currency gambling website which accepts Bitcoin
(BTC) as wagers.
It provides a Satoshi Dice game {\it Coinroll} (F),
where players can either wager on the 8 preset odds listed by the website,
or choose an odds of their own.
When further analyzing the data, we find that a few players placed an unusual large
amount of bets, where the top player placed more than 11 million bets.
Although these large number of bets prove the heavy-tailed distribution of
the number of bets of individuals, we have doubts that these players
are playing for the purpose of gambling.
As we have pointed out,
all the games discussed in this paper have negative expected payoffs.
Indeed, prior studies have raised
suspicion about the use of crypo-currency gambling websites as a way for
money laundering~\cite{fiedler2013}.
We will therefore exclude from our analysis gamblers
who placed more than half a million bets.
For bets wagered on the preset odds, we find that
more than $57\%$ are placed under the odds $1.98$, and we use these bets
to analyze the wager distribution.
On the other hand, since player-selected odds show a broader spectrum regarding
the risk attitude of gamblers, we focus on the odds distribution of
the player-selected odds. 
As already mentioned, we will exclude the bets from those players who placed at least half a million bets from our odds
distribution analysis.

Although crypto-currency has gained decent popularity in the financial
and technological world, in this paper we still measure the wager/wealth
deposited in forms of crypto-currencies in US dollars, since the wagers
in skin gambling are measured in US dollars.
The historical daily price data of crypto-currencies (Bitcion, Ethereum,
Bitcoin Cash) are obtained from CoinDesk~\cite{coindesk} (for Bitcoin)
and CoinMetrics~\cite{coinmetrics} (for Ethereum and Bitcoin Cash).

\subsection*{Ethics for data analysis}
The data collected and analyzed in this paper are all publicly accessible
on the internet, and we collect the data either with the consent of the
website administrators or without violating the terms of service or
acceptance usage listed on the hosting website.
The data we use do not include any personally identifiable information (PII), and we
further anonymize account-related information before storing them into our databases
to preserve players' privacy. In addition,
our data collection and analysis procedures are performed solely passively,
with absolutely no interaction with any human subject.
To avoid abusing the hosting websites (i.e., the gambling websites),
the request rates of data-collecting are limited to $1$ request per second.
Considering the legal concerns and potential negative effects of
online gambling~\cite{Martinelli2017, Millar2018, Kairouz2012, Gonzalez2017,
Gainsbury2015, Banks2017, Redondo2015, Macey2018}, our analysis
aims only to help better prevent adolescent gambling and problem gambling.

\subsection*{Parameter estimation and model selection}
In our analysis, the parameters of different distribution models are obtained
by applying Maximum Likelihood Estimation (MLE) \cite{Bauke2007}.
To select the best-fit distribution, we compare the models'
Akaike weights \cite{Burnham2002} derived from Akaike Information Criterion (AIC).
Note that analyzing the fitting results, we constantly found that
players show a tendency of using simple numbers
when allowed to place wagers with arbitrary amounts of virtual currency.
As a result, the curves of probability distribution functions
appear to peak at simple numbers, and the corresponding
cumulative distribution function shows a stepped behavior.
This makes the fitting more difficult, especially for the determination
of the start of the tail. To address this issue,
we choose the start of the tail $x_{min}$ such that we obtain a small
Kolmogorov\textendash Smirnov (K\textendash S) distance between
the empirical distribution and the fitting distribution, while maintaining
a good absolute fit between the complementary cumulative distribution functions
(CCDF) of the empirical distribution and the best-fitted distribution.
Candidate models for model selection in this paper include
exponential distribution, power-law distribution, log-normal distribution,
power-law distribution with sharp truncation,
power-law distribution with exponential cutoff, and
pairwise power-law distribution.
More details about parameter fitting and model selection can be found in the article
by Clauset et al. article~\cite{Clauset2009} as well as in the previous paper by the authors \cite{Wang2018}.

\section*{Data availability}
The datasets generated and/or analysed during the current study are available from the authors on reasonable request.





\section*{Acknowledgements}

This work is supported by the US National Science Foundation through grant DMR-1606814.

\section*{Author contributions statement}

X.W. and M.P. conceived the study,  X.W. wrote the computer codes and conducted the data analysis, X.W. and M.P. discussed the results and wrote the manuscript.

\section*{Additional information}

\textbf{Competing Interests}\\
The authors declare no competing interests.

\end{document}